\documentclass[11pt,letterpaper]{article}
\usepackage{jheppub}
\usepackage[english]{babel}
\usepackage{amssymb,amsmath}
\usepackage{mathtools}
\usepackage{mathrsfs}
\usepackage{bbold}
\usepackage{hyperref}
\usepackage{empheq}
\usepackage{subfigure}
\usepackage{xfrac} % Nice fractions
\usepackage{graphics,color}
\usepackage{slashed}
\usepackage{graphicx}
\usepackage{braket}
\usepackage{float}
\usepackage{tcolorbox}
\usepackage{booktabs}
\usepackage{cancel}

\newcommand{\be}{ \begin{equation}}
\newcommand{\ee}{ \end{equation}}

\title{Higgs Squared}

\author{Csaba Cs\'aki,}
\author{Ameen Ismail,}
\author{Maximilian Ruhdorfer,}
\author{and Joseph Tooby-Smith}

\emailAdd{csaki@cornell.edu}
\emailAdd{ai279@cornell.edu}
\emailAdd{m.ruhdorfer@cornell.edu}
\emailAdd{j.tooby-smith@cornell.edu}

\affiliation{Laboratory for Elementary Particle Physics, Cornell University, Ithaca, NY 14853, USA}

\abstract{We present a novel construction for a Higgs-VEV sensitive (HVS) operator, which can be used as a trigger operator in cosmic selection models for the electroweak hierarchy problem. Our operator does not contain any degrees of freedom charged under the SM gauge symmetries, leading to reduced tuning in the resulting models. Our construction is based on the extension of a two Higgs doublet model (2HDM) with a softly broken approximate global $D_8$ symmetry (the symmetry group of a square). A cosmic crunching model based on our extended Higgs sector has only  a percent level tuning corresponding to the usual little hierarchy problem. In large regions of parameter space the 2HDM is naturally pushed towards the alignment limit. A complete model requires the introduction of fermionic top partners to ensure the approximate $D_8$ symmetry in the fermion sector. We also show that the same extended Higgs sector can be used for a novel implementation of the seesaw mechanism of neutrino masses.  

}

\begin{document}

\maketitle	

%%%%%%%%%%%%%%%%%%%% 
\section{Introduction}
%%%%%%%%%%%%%%%%%%%%
%

Since the discovery of the 125 GeV Higgs ten years ago, all of its measured properties have been consistent with the Standard Model (SM). Yet, there are several theoretical and experimental motivations for considering models of new physics with extended Higgs sectors. These motivations include, among others, the Higgs naturalness problem (a.k.a. the hierarchy problem) and the explanation of neutrino masses.

Traditional approaches to the hierarchy problem use symmetries to protect the mass of the Higgs (for example, weak-scale supersymmetry), and typically predict new colored states around the TeV scale. Consequently, these models face pressure from the lack of discovery of beyond the SM particles at the LHC. This has motivated the development of new paradigms for addressing Higgs naturalness; in particular, a number of cosmological approaches to the hierarchy problem have been proposed in recent years~\cite{Graham:2015cka,Arkani-Hamed:2016rle,Geller:2018xvz,Cheung:2018xnu,Giudice:2019iwl,Strumia:2020bdy,Csaki:2020zqz,Arkani-Hamed:2020yna,TitoDAgnolo:2021pjo,TitoDAgnolo:2021nhd}. In these models the quadratically divergent corrections to the Higgs mass are unsuppressed. Instead some novel dynamics selects a small Higgs vacuum expectation value (VEV) in the early universe.

A generic feature of cosmological naturalness models is the presence of a ``trigger operator'', an operator which is sensitive to the Higgs VEV~\cite{TitoDAgnolo:2021pjo}. The trigger operator couples to new physics, and induces (or ``triggers'') a cosmological event, such as a phase transition, preventing large Higgs VEVs. Common trigger operators in the literature include $\lvert H \rvert^2$, $\mathrm{tr} \: G \tilde{G}$ (where $G$ is the gluon field strength), and in the context of the two Higgs doublet model (2HDM), $\Phi_1^\dagger \Phi_2$.

The presence of an operator that acquires a VEV which is sensitive to the Higgs VEV in some useful way is not unique to cosmological naturalness models. Another example occurs in neutrino physics, in particular the type II seesaw mechanism. Here a new scalar field $\Delta$ is introduced to the SM which couples to the left-handed leptons as $\overline{L}^c \Delta L$, and is therefore charged under the SM gauge group. The scalar potential of this model has a term $H\Delta H$, which makes $\Delta$ sensitive to the Higgs VEV: $\langle \Delta \rangle \sim \langle \lvert H \rvert^2 \rangle / m_{\Delta}$. The mass of $\Delta$ is assumed to be much larger than the electroweak scale, which means that the VEV of $\Delta$ is also suppressed relative to the Higgs VEV, and this feature is what accounts for the small neutrino mass.

Let us refer to this broad class of operators as Higgs VEV-sensitive (HVS) operators.
The HVS operators introduced above have one thing in common: they are composed of fields charged under the SM gauge group. In certain situations this is undesirable.  For example, in Ref.~\cite{Csaki:2020zqz}, the choice of the $\lvert H \rvert^2$ trigger leads to electroweak gauge boson partners; some degree of fine-tuning is then required to push the mass of these states up to the TeV scale to avoid experimental bounds. Furthermore, this dependence on the SM gauge group limits the application of such operators, due to the often-strong constraint of gauge invariance.

The aim of this paper is to build a trigger/HVS operator entirely of SM singlets which nevertheless tracks the Higgs VEV. Our construction will be based on a 2HDM with additional discrete symmetries. The key role will be played by the finite group $D_8$, corresponding to the symmetries of a square (hence the title Higgs squared). This will allow us to introduce the SM singlet field $B$, which will be the trigger/HVS operator that we are after. We will show that $B$ acquires a VEV of the form $\langle B\rangle \propto v_1 v_2/\Lambda$ where $v_{1},v_2$ are the VEVs of the two Higgs doublets and $\Lambda$ is the UV cutoff, while having a mass of $\mathcal{O}(\Lambda)$. The $D_8$ symmetry serves two purposes in our model: i) it forbids terms in the scalar potential which would prevent the sensitivity of $\langle B \rangle$ to the Higgs VEVs and ii) if not explicitly broken and $v_1, v_2 \neq 0$, it leads to $v_1^2 = v_2^2 = v^2/2$. In combination this will allow us to cap the Higgs VEV $v^2 = v_1^2 + v_2^2$ by imposing an external (cosmological) constraint on the VEV of $B$, our trigger operator. Note that~\cite{Draper:2016cag} also considered a similar finite symmetry in the context of a 2HDM in order to address ii) and obtain a partially natural 2HDM. Our model is distinguished by the presence of the $B$ scalar which allows us to extend the partial to full naturalness.

The $D_8$ symmetry has some interesting phenomenological consequences. Foremost it requires fermionic partners for the SM fermions in order to construct $D_8$-invariant Yukawa couplings. However, an exact $D_8$ symmetry would predict degenerate Higgs bosons and fermionic partners degenerate with the SM fermions. Thus we allow the $D_8$ symmetry to be softly broken by vector-like masses of the fermionic partners and a small difference in the Higgs mass parameters $\mu^2 = m_{\Phi_1}^2-m_{\Phi_2}^2\ll \Lambda$. Experimental constraints force the fermionic partners to be heavier than $\mathcal{O}({\rm TeV})$ (see Section~\ref{sec:TPpheno}) which introduces the little hierarchy into our model, corresponding to a percent-level tuning.

At low energies the phenomenology of our model is that of a 2HDM with 2 CP-even Higgses, one CP-odd Higgs and one charged Higgs --- all except one CP-even Higgs having electroweak-scale masses --- with the addition of TeV-scale fermionic partners. In particular the low-energy phenomenology is insensitive to the mechanism that ensures a small $B$ VEV. 
However, the selection of a small $B$ VEV forces us into particular corners of the 2HDM parameter space which naturally splits into two regions, characterized by the mass of the non-SM-like CP-even Higgs. In the first region, the CP-even Higgs is much lighter than the SM-like Higgs, with a mass as small as $\mathcal{O}(100{\rm ~MeV})$. This light Higgs is long-lived, and its phenomenology is similar to a scalar that mixes weakly with the SM Higgs. Interestingly, in this regime the 2HDM is naturally pushed toward the so-called ``alignment limit'', in which the two CP-even Higgses align ``parallel'' and ``perpendicular'' to the direction of the electroweak VEV. Any phenomenologically viable 2HDM needs to be close to the alignment limit and our model achieves this naturally in this region. For other examples in the literature of naturally aligned 2HDMs, see Refs.~\cite{BhupalDev:2014bir,Pilaftsis:2016erj,Draper:2016cag,Draper:2020tyq,Haber:2022swy}.

In the second region of parameter space, the second CP-even Higgs has a mass of the same order as the SM-like Higgs, and it may be the heavier or the lighter of the two. The phenomenology is essentially that of a generic 2HDM in this regime. The electroweak scale is natural, but the alignment limit is not, such that some amount of tuning is required to reach the proximity of the alignment limit.

The paper is organized as follows. We start in Section~\ref{sec:overview} with a high-level overview of how our setup solves the hierarchy problem. We also discuss an explicit realization of a cosmological selection mechanism of a small $B$ VEV, based on a modification of the crunching mechanism previously employed to address fine-tuning problems in Refs.~\cite{Csaki:2020zqz,Bloch:2019bvc}. 
Following this qualitative overview we properly introduce our 2HDM extended by the $B$ scalar in Section~\ref{sec:model} and demonstrate that its VEV scales as $\langle B \rangle \sim v^2/\Lambda$. This sets the stage for the exploration of its phenomenology in Section~\ref{sec:pheno}. In Section~\ref{sec:Neutrinos} we outline a second application of our mechanism: generating small neutrino masses through a modification of the type II seesaw mechanism. We introduce a complex scalar field $\Delta$ with a mass of order $\Lambda$. The symmetries of our model forbid a dimension-4 interaction of $\Delta$ to left-handed leptons but allow for a dimension-5 one involving $B$. The $B$ scalar leads to a suppression of the neutrino mass  and allows $\Lambda$ to be as low as $10^3~\mathrm{GeV}$, in contrast to the $\Lambda\sim 10^{14}~\mathrm{GeV}$ typical of seesaw models.

%
%%%%%%%%%%%%%%%%%%%% 
\section{Overview} \label{sec:overview}
%%%%%%%%%%%%%%%%%%%%
%
In this section we give an overview of our model in the context of the hierarchy problem. As we will explain in Section~\ref{sec:Neutrinos}, the model can be used for neutrino physics as well. At the heart of our model is a SM singlet, real scalar field $B$. We will assume that a mechanism auxiliary to our model constrains the VEV of $B$ to lie in a finite range, $0 < | \langle B \rangle | < B_{\rm crit}$, where $B_{\rm crit}$ may be exponentially smaller than the cutoff of the theory without fine-tuning. We will shortly discuss one possible such mechanism which adapts the crunching dynamics introduced in Ref.~\cite{Csaki:2020zqz}.

In order to communicate the low scale $B_{\rm crit}$ to the SM and use it to solve the hierarchy problem, we want $B$ to track the Higgs VEV in such a way that the $0< | \langle B \rangle | < B_{\rm crit}$ regime corresponds to Higgs VEVs of order TeV or smaller. This can be achieved through a trilinear coupling of $B$ and two Higgs fields, schematically $HHB$, if in addition the tadpole term $B$ is suppressed or absent. There are two simple ways to do this: either we interpret $B$ as a pNGB, which would require its potential to be suppressed by the spurion that breaks the Goldstone symmetry, or we introduce some symmetry which forbids the tadpole term altogether. We follow the latter route. However, this route is infeasible with only one Higgs doublet, since a coupling of $B$ to $H^\dagger H$ would require $B$ to be a singlet under all symmetries, while coupling $B$ to $HH$ is not possible if $B$ is an SM gauge group singlet. To overcome this we will consider a 2HDM with two Higgses $\Phi_1$ and $\Phi_2$.

By introducing a $\mathbb{Z}_2$ symmetry under which $B$ and one of the Higgs doublets $\Phi_2$ are odd we can allow the term 
\begin{align}
c_{B\Phi}\Lambda B(\Phi_1^\dagger\Phi_2+\Phi_2^\dagger \Phi_1),
\end{align}
while forbidding the tadpole term of $B$. To see how this helps us, consider the quadratic terms of the Higgses and $B$,
\begin{align}\label{eq:2HDMmasses}
	V\supset c_H\Lambda^2(\Phi_1^\dagger\Phi_1+\Phi_2^\dagger\Phi_2)+\mu^2 (\Phi_1^\dagger\Phi_1-\Phi_2^\dagger\Phi_2)+c_B \Lambda^2B^2.
\end{align}
We will assume that the parameter $c_H$ is cosmologically scanned, and that $c_B$ is positive. In a 2HDM to solve the hierarchy problem both Higgs VEVs must be $\ll \Lambda$. Further, in order for $B$ to have a non-zero VEV, they must both be non-zero. If $\mu^2$ is of order $\Lambda^2$ then these two conditions cannot be satisfied simultaneously. In order to achieve $\mu^2\ll \Lambda^2$ we need another symmetry that has $\mathbb{Z}_2$ as a subgroup and allows $c_H, c_{B\Phi}, c_B$ but forbids $\mu^2$, such that $\mu^2$ is a small, technically natural explicit breaking of this symmetry. The simplest choice is to use a $D_8$ nonabelian discrete symmetry, the symmetry group of a square. The same symmetry was studied in the absence of $B$  and in the context of the hierarchy problem in~\cite{Draper:2016cag}. 

The VEV of $B$ takes the schematic form $\langle B\rangle\sim v_1v_2/\Lambda$. As explained we assume that the hidden sector constrains the size of the VEV of $B$. If $v_1\sim v_2\sim v$  (where $v^2=v_1^2+v_2^2$) this directly constrains the VEV of the Higgses, which will then naturally be much smaller than $\Lambda$ (case 1). If however, $v_2 \ll v_1$  (say), the constraint on $\langle B \rangle$ does not directly relate to a bound on $v^2$. But Eq.~\eqref{eq:2HDMmasses} shows that $v_2^2/v_1^2 \sim (c_H \Lambda^2 - \mu^2)/(c_H \Lambda^2 + \mu^2)$, assuming $v_1,v_2 \neq 0$. This tells us that $v_2 \ll v_1$ can only occur if $c_H\sim \mu^2/\Lambda^2$. Thus in this case, we are pushed to a value of the Higgs VEV of $v^2\sim \mu^2$ which is again much smaller than $\Lambda$ (case 2). Therefore, both of these cases solve the hierarchy problem. We summarize them as
\begin{equation}\label{eq:hierarchycases}
    \tan \beta \sim 1, v \ll \Lambda\quad  {\rm ~(case~1)}, \qquad \tan \beta \gg 1, v \sim \mu\quad {\rm ~(case~2)}
\end{equation}
where $\tan \beta = v_2 / v_1$.

To understand when each case occurs in our model, first note there exists a single minimum of our scalar potential with $\langle B \rangle \neq 0$ and consequently also $v_1, v_2 \neq 0$. There is a critical point where $v_1 , \langle B \rangle = 0$, which occurs at a characteristic value of the Higgs mass parameter $c_{H,0} \sim \mu^2/\Lambda^2$. It is convenient to define a dimensionless parameter $r$, which is linear in $c_H$ and vanishes at the critical point:
\begin{equation}
    r = \frac{c_H - c_{H,0}}{c_{H,0}}.
\end{equation}
The VEV $\langle B \rangle$ is roughly related to $r$ as
\begin{equation}\label{eq:Bvevscaling}
    \langle B \rangle^2 \sim \frac{ \mu^4}{\Lambda^2 }r .
\end{equation}
Note that $\langle B \rangle^2$ increases monotonically with $r$ and vanishes when $r = 0$. We will show that the Higgs VEVs scale as
\begin{equation}\label{eq:Higgsvevscaling}
    v^2\sim \mu^2(1+r), \quad \tan\beta \sim \frac{2}{r}+1 .
\end{equation}
From Eq.~\eqref{eq:Higgsvevscaling}, we see that $v^2$ also increases monotonically with $r$. Therefore $B$ is sensitive to the Higgs VEV. As we approach the critical point $r \rightarrow 0$, $v^2 \sim \mu^2$ and $v_1 \rightarrow 0$.

Specifically, when $r\ll 1$ (equivalently $|\langle B \rangle | \ll \mu^2/\Lambda$), we have $v_2\gg v_1$ and $v^2\sim \mu^2$. This corresponds to our second case that solves the hierarchy problem in Eq.~\eqref{eq:hierarchycases}. Further away from the critical point, for $r\gtrsim 1$, we have $\tan\beta\sim 1$ and $v^2\sim \mu^2 r$. As long as $r\ll \Lambda^2/\mu^2$ it follows that $v^2\ll \Lambda^2$, corresponding to the first case in Eq.~\eqref{eq:hierarchycases}.  

For this mechanism to work it is critical that the $D_8$ symmetry is only softly broken. This forces us to introduce fermionic partners for the SM fermions to construct $D_8$ invariant Yukawa couplings. We softly break the $D_8$ symmetry in the fermion sector through TeV scale vector-like mass terms for the fermionic partners. We further assume that $\mu^2$, the soft-breaking parameter in the scalar potential, is loop suppressed compared to the vector-like mass terms of the fermionic partners. Thus to prevent a radiatively unstable hierarchy between the two soft-breaking parameters the fermionic partner mass and $\mu^2$, whose natural value is a loop factor below the fermion partner mass, we concentrate on the region $r\lesssim 1$ for which this tuning is minimized.

\begin{figure}[t]
\begin{center}
\includegraphics[page=1,width=3in]{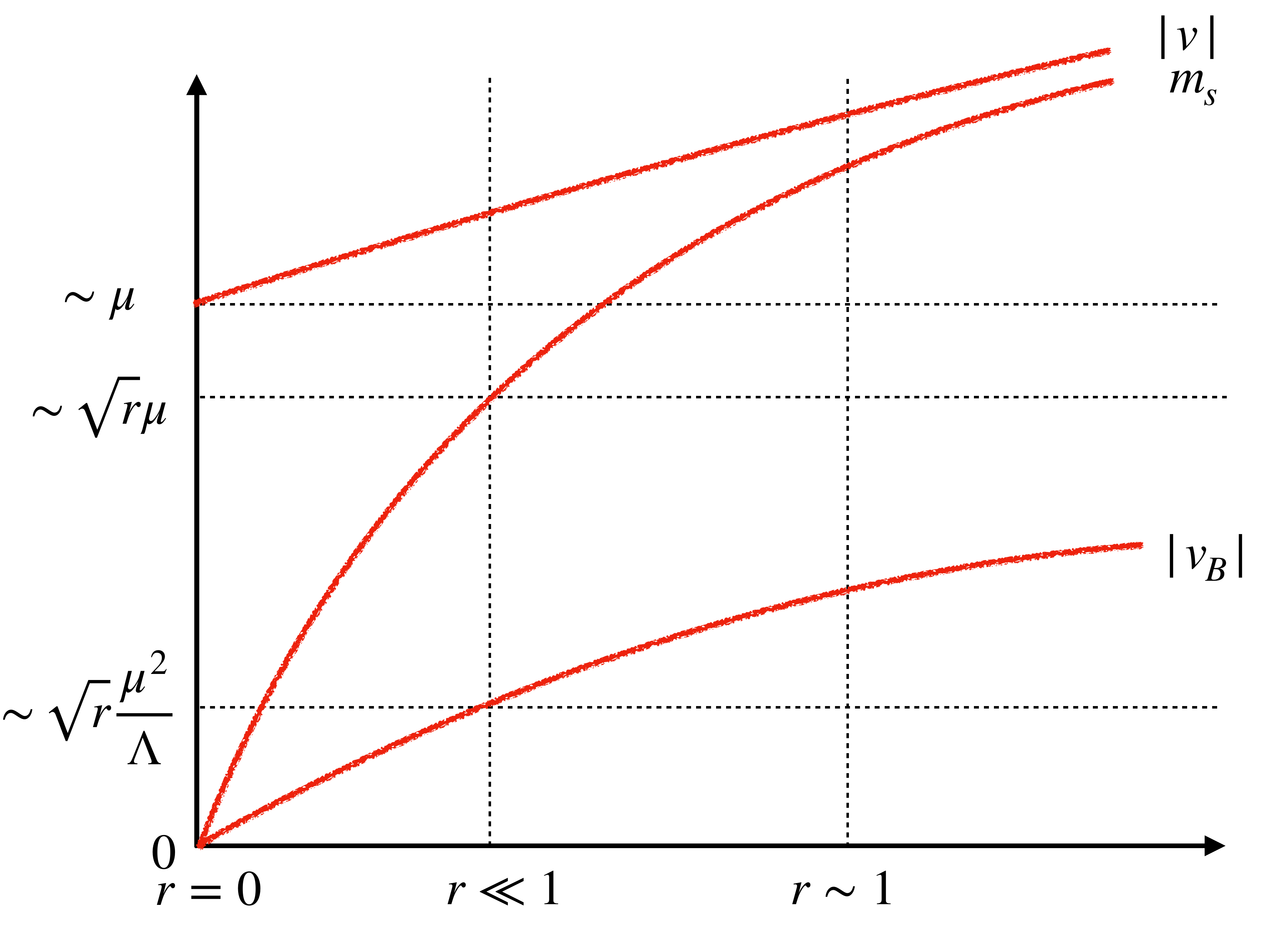}
\includegraphics[page=2,width=3in]{./Figures/Quantity_Sketch_V3.pdf}
\end{center}
\caption{An illustration of how our physical quantities change in the different $r$ regions. The x-axes start at the point corresponding to $r=0$, and the dashed lines give the rough scale of the parameters at a fixed $r$. The left panel shows quantities of mass dimension $1$. The right panel shows quantities of mass dimension $0$.}
\label{fig:Sketch_of_parameters}
\end{figure}

The upshot is that we solve the hierarchy problem when $r \lesssim 1$, corresponding to $|\langle B \rangle | \lesssim \mu^2/\Lambda$. We obtain a naturally small electroweak scale of order $\mu$. After electroweak symmetry breaking, we will have five physical scalar fields. The one associated with $B$ sits roughly at the UV scale $\Lambda$. The others comprise the usual 2HDM fields: two CP-even Higgses which we denote $s$ and $h$ (contrary to normal 2HDM convention), a CP-odd Higgs $A$, and a charged Higgs $H^\pm$. Their masses are
\begin{align}
	m_s^2\sim \mu^2 r, \quad m_h^2, m_{H^\pm}^2, m_A^2\sim \mu^2(1+r).
\end{align}
The mass of $h$ sits at the same scale as $v$. Thus we can interpret it as the SM-like Higgs. In the region $r\ll 1$, the mass of $s$ is generically much smaller than the EW scale and its couplings to the SM (proportional to the alignment parameter $\sin(\beta-\alpha)$) are suppressed, leading to interesting phenomenology. The scaling of the model parameters with $r$ is summarized in Figure~\ref{fig:Sketch_of_parameters}.

\subsection{The hidden sector}
To solve the hierarchy problem some hidden sector dynamics must select a small VEV for $B$.
One way to do this, although certainly not the only way, is through the crunching mechanism introduced in Ref.~\cite{Csaki:2020zqz}. We briefly describe the mechanism here; a self-contained, more detailed description is provided in Appendix~\ref{sec:crunchingappendix}. The way in which one obtains a small $\langle B \rangle$ does not affect the phenomenology we discuss in Section~\ref{sec:pheno}.

We postulate a multiverse of causally disconnected patches wherein a scanning sector sets the Higgs mass-squared parameter, $c_H$, and thereby $r$ and $\langle B \rangle$, in each patch, up to some cutoff scale. We introduce a spontaneously broken conformal sector that couples to the scalar singlet operator $B^2$, which now justifiably can be called a ``trigger operator''. In the 5D dual description of the conformal field theory (CFT), this means $B$ propagates in the AdS bulk, while the remaining SM fields are localized on the UV brane. The interaction between $B$ and the dilaton (the Goldstone boson of spontaneously broken scale invariance) causes each patch to rapidly undergo a cosmological crunch unless the VEV of $B$ lies in a finite range,
\begin{equation}\label{eq:finiteBrange}
0<B_{\emptyset}\le |\langle B \rangle|\le B_{\rm crit} .
\end{equation}
Essentially, the dilaton potential is sensitive to $\langle B \rangle$. A value of $\langle B \rangle$ larger than $B_{\rm crit}$ triggers a phase transition in the dilaton, leading to a vacuum with a large negative cosmological constant which causes a crunch. The only cosmologically long-lived patches are those where $\langle B \rangle$ lies inside this range. Thanks to the conformal symmetry, we can naturally have $B_{\rm crit} \ll \Lambda$. The crunching sector also generates (for example, through the addition of a bulk confining gauge group) a minimal VEV $B_{\emptyset}$ such that patches where $| \langle B \rangle | < B_{\emptyset}$ crunch, removing small or vanishing Higgs VEVs from the landscape. An explicit form for the dilaton potential is given in Apppendix~\ref{sec:crunchingappendix}.

Because $B$ is an SM singlet, our model does not suffer from the phenomenological drawbacks of Ref.~\cite{Csaki:2020zqz} --- namely, electroweak-scale Kaluza--Klein partners of the electroweak gauge bosons, resulting in a little hierarchy and an $\mathcal{O}(10^{-4}$--$10^{-3})$ tuning in the crunching sector. We do not have Kaluza--Klein modes of any SM fields, and thus no little hierarchy nor tuning in the crunching sector. In addition, the phenomenology of our model is very different from Ref.~\cite{Csaki:2020zqz}. Our main experimental signatures are those of a 2HDM, rather than a GeV-scale dilaton like in Ref.~\cite{Csaki:2020zqz}.

We emphasize again that a different mechanism could be used to select a small $B$ VEV. Moreover, the 2HDM phenomenology is totally independent of the crunching mechanism.

%%%%%%%%%%%%%%%%%%%% 
\section{The Model} \label{sec:model}
%%%%%%%%%%%%%%%%%%%%
With the above motivation in mind, let us now construct the full model.

%%%%%%%%%%%%%%%%%%%%
\subsection{The scalar potential} \label{sec:ScalarPotential}
%%%%%%%%%%%%%%%%%%%%
%
\begin{table}
\begin{center}
\renewcommand\arraystretch{1.2}
\begin{tabular}{|c|c|c|c|c|c|}
\hline
\S & Symbol & Type & SM-rep & $D_8$ irrep & $\mathbb{Z}_2\times \mathbb{Z}_2$ irrep\\ \hline
 \ref{sec:ScalarPotential}& $\Phi$ & Complex scalar & $(\mathbf{1,2})_{1/2}$ & $r_2$ & $\Phi_1\sim(-,+)$ \\
& & & & & $\Phi_2 \sim(+,-)$  \\ \hline
\ref{sec:ScalarPotential} & $B$ &Real scalar & $(\mathbf{1,1})_0$ & $r_{--}$ & $(-,-)$ \\
\hline
\ref{sec:fermionSector}& $Q_3$ &LH Weyl & $(\mathbf{3,2})_{1/6}$& $r_{++}$ & $(+,+)$ \\ \hline
\ref{sec:fermionSector}& $\tilde T_R$ & RH Weyl & $(\mathbf{3,1})_{2/3}$ & $r_2$ & $\tilde T_{1R}\sim (-,+)$ \\
& & & & & $\tilde t_R\sim(+,-)$ \\ \hline
\ref{sec:fermionSector}& $\tilde T_L$ & LH Weyl &  $(\mathbf{3,1})_{2/3}$ & $r_{++}$ & $(+,+)$ \\
\hline \hline
\ref{sec:Neutrinos}& $L$ & LH Weyl & $(\mathbf{1,2})_{-1/2}$ & $r_{++}$ & $(+,+)$ \\ \hline
\ref{sec:Neutrinos}& $\Delta$ & Complex scalar & $(\mathbf{1,3})_1$ & $r_{+-}$ & $(-,-)$ \\ \hline
\end{tabular}
\end{center}
\caption{A summary of the fields introduced to our model and its applications. The fields below the double horizontal lines are of relevance to the neutrino model in Section~\ref{sec:Neutrinos} only, and can otherwise be ignored.}
\end{table}

 The model is based on a 2HDM with a $\mathbb{Z}_2$ symmetry. We have two complex scalar fields $\Phi_1$, and $\Phi_2$ carrying the Higgs representation, $(\mathbf{1},\mathbf{2})_{1/2}$,  of  $SU(3)\times SU(2)\times U(1)$. We take $\Phi_1$ to be even and $\Phi_2$ to be odd under the $\mathbb{Z}_2$ symmetery. 
In addition, we introduce a real scalar field $B$ which is a singlet under the SM gauge group, and is odd under $\mathbb{Z}_2$. On top of the $\mathbb{Z}_2$ symmetry we include a CP symmetry under which $\Phi_i\mapsto \Phi_i^\ast$, and $B\mapsto B$. 

The most general scalar potential invariant under the full symmetry is given by 
\begin{align} \label{eq:2HDMPotential}
V(\Phi_1,\Phi_2,B)=V_\Phi^\prime(\Phi_1,\Phi_2)+V_B(B)+V_{\Phi B} (\Phi_1,\Phi_2,B).
\end{align}
where
\begin{align}
V_\Phi^\prime(\Phi_1,\Phi_2)&=c_H\Lambda^2(\Phi_1^\dagger\Phi_1+\Phi_2^\dagger\Phi_2)+\mu^2 (\Phi_1^\dagger\Phi_1-\Phi_2^\dagger\Phi_2)+\frac{1}{2}\lambda_1^\prime (\Phi_1^\dagger\Phi_1)^2+\frac{1}{2}\lambda_2^\prime(\Phi_2^\dagger\Phi_2)^2 \label{eq:VphiPrime}\nonumber\\&+\lambda_3^\prime(\Phi_1^\dagger\Phi_1)(\Phi_2^\dagger\Phi_2)+\lambda_4^\prime (\Phi_1^\dagger\Phi_2)(\Phi_2^\dagger\Phi_1)+\frac{1}{2}\lambda_5^\prime((\Phi_1^\dagger\Phi_2)^2+(\Phi_2^\dagger\Phi_1)^2)
\\
V_B(B)&=\frac{1}{2}c_B \Lambda^2  B^2+\frac{1}{4}\lambda_B B^4 \label{eq:VB}
\\
V_{\Phi B}(\Phi_1,\Phi_2,B)&=c_{B\Phi}\Lambda B(\Phi_1^\dagger\Phi_2+\Phi_2^\dagger \Phi_1)+\lambda_{1 B} B^2\Phi^\dagger_1 \Phi_1+\lambda_{2B} B^2\Phi^\dagger_2 \Phi_2 \label{eq:BPotential}
\end{align}
and all parameters are real. {We assume that all mass scales which are not protected by symmetries (see Section~\ref{sec:TechnicalNatural}) are of the same order as the UV cutoff $\Lambda$ with order-one coefficients $c_i$.} The reason for the primes in Eq.~\eqref{eq:VphiPrime} will become apparent in Section~\ref{subsec:PotentialMinimization}. 

%%%%%%%%%%%%%%%%%%%% 
\subsection{Technical naturalness and accidental symmetries} \label{sec:TechnicalNatural}
%%%%%%%%%%%%%%%%%%%% 
When $\mu=0$, $\lambda_1^\prime=\lambda_2^\prime$, and $\lambda_{1B}=\lambda_{2B}$ the scalar potential has an enhanced $D_8$ symmetry.\footnote{Note that in the 2HDM literature the group $D_8$ is often replaced with a $\mathbb{Z}_2\times \mathbb{Z}_2$ symmetry, and the representation $r_2$ of $D_8$ with a projective representation of $\mathbb{Z}_2\times \mathbb{Z}_2$; see e.g.~\cite{Haber:2021zva}.} This makes it technically natural to have $\mu \ll \Lambda$. 

The group $D_8$ is one of the lowest-order nonabelian  groups and corresponds to the symmetry group of a square. This order-8 group is generated by a $90^\circ$ rotation $a$ and reflection $x$ along the horizontal axis. Put formally, it has the presentation 
\begin{align}
D_8=\langle x,a\mid a^4=x^2=xaxa =e\rangle,
\end{align}
where $e$ is the identity. The first condition tells us that four  $90^\circ$ rotations is equal to the identity, the second that two reflections  is equal to the identity, and the last that a  $90^\circ$ rotation followed by a reflection, repeated twice, is equal to the identity.
\begin{table}
\begin{equation*}
	\begin{array}{|c|cc|c|c|}
	\hline
		\text{Irrep} & a & x & \text{$\{e,a^2,x,a^2x\}\cong \mathbb{Z}_2\times\mathbb{Z}_2 $ decomp.} & \text{$\{e,x\}\cong \mathbb{Z}_2 $  decomp.}\\ \hline
		r_{++} & +1 & +1 & (+,+) &+ \\
		r_{-+} & -1 & +1 &  (+,+) &+ \\
		r_{+-} & +1 &-1 &  (-,-) &- \\
		r_{--} & -1 & -1 &  (-,-)&- \\
		r_2 &\begin{pmatrix} 0 & -1 \\ 1 & 0 \end{pmatrix} &\begin{pmatrix} 1 &0 \\ 0 & -1 \end{pmatrix}&(+,-)\oplus( - ,+)& - \oplus +\\ \hline
	\end{array}
\end{equation*}
\caption{The irreducible representations of $D_8$, and their decompositions under relevant subgroups. \label{tab:IrrepsD8}}
\end{table}
Since $D_8$ is non-abelian it has at least one irrep which is not one-dimensional. In fact it has exactly one, which is two-dimensional. All of the irreps of $D_8$ are summarized in Table~\ref{tab:IrrepsD8}.

In the $D_8$ symmetric limit of the scalar potential the two Higgs doublets $\Phi_1$ and $\Phi_2$ can be combined into the 2D irrep $r_2$, while $B$ sits in the 1D irrep we denote $r_{--}$. Here the subscript ``$--$'' tells us the representation of $a$ and $x$, respectively. 

The $\mathbb{Z}_2$ symmetry group of our model corresponds to the sugbroup of $D_8$ consisting of reflections $\{ e, x\}$. However, looking at the potential in Eqs.~\eqref{eq:2HDMPotential}--\eqref{eq:BPotential} one can easily deduce that it is strictly invariant under the larger subgroup  of $D_8$ generated by reflection and $180^\circ$ rotation (e.g. $a^2 x$ and $x$). This subgroup is isomorphic to $ \mathbb{Z}_2 \times \mathbb{Z}_2$, with $\Phi_1$ transforming in the $(-,+)$ irrep, $\Phi_2$ transforming in the $(+,-)$ irrep and $B$ transforming in the $(-,-)$ irrep.

In the following we will assume that our model possesses this approximate $D_8$ symmetry which is only softly broken in the fermion sector through TeV scale vector-like masses for fermionic partners of the SM fermions which we introduce in Section~\ref{sec:fermionSector}. This breaking is communicated to the scalar sector via loops of SM fermions and their partners, with the dominant contribution originating from the top quark. Thus $\mu^2$ is generated radiatively and neglecting $\mathcal{O}(1)$ numbers it is therefore naturally of the size
\begin{equation}
    \mu^2 \sim \frac{y_t^2 M_{\tilde T}^2}{16 \pi^2} \log \frac{ \Lambda }{M_{\tilde T}}\,,
\end{equation}
where $y_t$ is the top Yukawa coupling and $M_{\tilde T}$ is the bare vector-like mass parameter for the top partner. The phenomenological requirement that $\mu^2 \ll y_t^2 M_{\tilde T}^2/(16\pi^2)$ introduces a little hierarchy into our model. See Section~\ref{sec:Tuning} for a discussion about the required tuning.

%%%%%%%%%%%%%%%%%%%
\subsection{Minimizing the potential}
\label{subsec:PotentialMinimization}
%%%%%%%%%%%%%%%%%%%
In this section we want to study the analytic minimization of the scalar potential. To do this, we will make some generic assumptions about the model parameters which we give in Appendix~\ref{sec:workingassumptions}. Among others these include conditions to ensure the boundedness of the potential and stability of minima with $\langle B \rangle\ne 0$ and $\langle B \rangle\ll \Lambda$, which will be our focus in the following. Via the assumptions made in Appendix~\ref{sec:workingassumptions}, such a minimum is also CP- and charge-conserving. The symmetry of our model allows us to rotate the Higgs VEVs into the form
\begin{align}\label{eq:Form_of_Phi}
\langle \Phi_1\rangle=\frac{v}{\sqrt{2}} \begin{pmatrix} 0 \\ \cos\beta\end{pmatrix},\quad \langle \Phi_2 \rangle =\frac{v}{\sqrt{2}} \begin{pmatrix} 0\\ \sin\beta\end{pmatrix}.
\end{align}
Given this alone, it is not possible to find the minima of $V$ analytically. Thus we note that when $\langle B \rangle\ll \Lambda$ and when $\lambda_{1B}-\lambda_{2B}$ is small (due to the approximate $D_8$ symmetry), then $V$ can be approximated by the same potential with $\lambda_B=0$ and $\lambda_{1B}=\lambda_{2B}=:\lambda_{\Phi B}$.
 
Even after this approximation is made, the minima cannot be found analytically. On solving the minimization equations, we can write $\langle B \rangle$ as a complicated function of the form
\begin{align}
\langle B \rangle^2=f(v^2, \Lambda,\mu, c_{B\Phi}, c_B,\lambda_i^\prime, \lambda_{B\Phi}).
\end{align} 
Although complicated, $f$ can be shown to have two useful properties which together allow us to call $B^2$ a HVS operator:
\begin{enumerate}
\item It increases monotonically in $v^2$ when all other inputs are fixed.
\item At $\langle B \rangle=0$ we have $v\sim \mu$.
\end{enumerate}
{The first result is in part due to our generic assumptions in Appendix~\ref{sec:workingassumptions} . The second result can be explained by the fact that when $\langle B \rangle$ is zero, the 2HDM must be in the inert phase, meaning the VEV lies in just one of $\Phi_1$ or $\Phi_2$. Since the only generic thing distinguishing  these Higgses in the potential is $\mu^2$, this occurs at $c_H\sim \mu^2/\Lambda^2$ and thus $v^2\sim \mu^2$. }

{A function $x^2=f(y^2)$ which increases monotonically as $y^2$ increases, and which is zero at  $y^2 \sim c$ for some $c$, has the property that being in a region close enough to $x=0$ implies that $y^2\sim c$. Applying this to our function above tells us that $\langle B \rangle$ close to zero implies $y^2\sim \mu^2\ll \Lambda^2$. If we assume that $|\langle B \rangle| \ll \mu$, then it is appropriate to neglect $\lambda_{1B}$ and $\lambda_{2B}$ entirely, whilst keeping the quadratic terms in $\Phi$. }
 {We will at this point also set $\lambda_1^\prime=\lambda_2^\prime=:\lambda^\prime$. The quantity $\lambda_1^\prime-\lambda_2^\prime$ is expected to receive radiative corrections logarithmic in the ratio of the top to top partner mass (which we will shortly introduce)~\cite{Draper:2016cag}. A detailed analysis shows that  setting  $\lambda_1^\prime=\lambda_2^\prime$ does not change the qualitative features of our model.  Full expressions for quantities found in the following when $\lambda_1^\prime\ne \lambda_2^\prime$  are given in Appendix~\ref{sec:workingassumptions}.  } Let us denote by $\tilde V$ the potential with $\lambda_{B},\lambda_{1B},\lambda_{2B}=0$ and $\lambda_1^\prime= \lambda_2^\prime$.

Finding the minima of $\tilde V$ is analytically possible since we can use the minimization condition for $B$ to replace it with 
\begin{align}\label{eq:Bmin}
B=-\frac{c_{B\Phi}}{c_B \Lambda} (\Phi^\dagger_1\Phi_2+\Phi_2^\dagger\Phi_1).
\end{align}
On substituting this into $\tilde V$, we get a potential $V_\Phi(\Phi_1,\Phi_2)$ of the same form as $V_\Phi^\prime(\Phi_1,\Phi_2)$, except with quartic coefficients $\lambda_i$ instead of $\lambda_i^\prime$, which are related to each other by
\begin{align}
\lambda^\prime=\lambda,\quad \lambda_{3}=\lambda_{3}^\prime, \quad \lambda_{4,5}=\lambda_{4,5}^\prime-\frac{c_{B\Phi}^2}{c_B}.
\end{align}
In what follows we will use the notation $\lambda_{345}:=\lambda_3+\lambda_4+\lambda_5$ and $\lambda_{45}:= \lambda_4+\lambda_5$.

The potential $V_\Phi(\Phi_1,\Phi_2)$ is simply that of a general 2HDM with a specific $\mathbb{Z}_2\times \mathbb{Z}_2$ symmetry, so we can use standard results to study it (see e.g.~\cite{Branco:2011iw,Haber:2015pua} and references therein). 
Recall we are interested in minima of the potential for which both $v_1 , v_2 \neq 0$ and consequently also $\langle B \rangle \neq 0$. There is one such minimum for our potential, with Higgs VEVs as given in Eq.~\eqref{eq:Form_of_Phi}. This exists for $c_H < c_{H,0}$ with 
\begin{equation}
	c_{H,0} = -\frac{\lambda + \lambda_{345}}{\lambda - \lambda_{345}} \frac{\mu^2}{\Lambda^2}\,.
\end{equation}
At the critical point, i.e. for $c_H = c_{H,0}$, it holds that $\langle B\rangle = v_1 =0$. For the following discussion it is convenient to introduce a dimensionless parameter $r$, which is linear in $c_H$
\begin{equation}\label{eq:rdefinition}
	r = \frac{c_H - c_{H,0}}{c_{H,0}}\,.
\end{equation}
The condition $c_H < c_{H,0}$ for the existence of the phase translates into $r > 0$, since $c_{H,0} < 0$. Expressing the Higgs VEVs in terms of $r$ yields
\begin{align}
\frac{v^2}{\mu^2}&=
\frac{4}{\lambda-\lambda_{345}}(1+r),\label{eq:HiggsVev}\\
\tan^2\beta&=
\frac{2}{r}+1\,. \label{eq:tanBeta}
\end{align}
(recall $\tan \beta = v_2/ v_1$). Using Eq.~\eqref{eq:Bmin} we can find an explicit expression for the VEV of $B$ 
\begin{align} \label{eq:vevB}
\frac{\Lambda}{\mu^2}\langle B \rangle=
-\frac{2c_{B\Phi}}{c_B(\lambda-\lambda_{345})}\sqrt{2r+r^2}\,.
\end{align}

{To remake a connection to the hierarchy problem, we observe that both $v^2$ and $|\langle B \rangle|$ increase monotonically with $r$ and for $r\rightarrow 0$  (i.e. close to the critical point) the VEV $\langle B \rangle\rightarrow 0$. This fact will be used to cosmologically ensure that we sit near the critical point, where as can be seen from Eq.~\eqref{eq:HiggsVev} $v^2\sim \mu^2(1+r)$ , and in particular $v^2\sim \mu^2$ in the case when $r \lesssim 1$.} 
\subsection{Mass eigenstates and spectrum}
To study the physical scalar sector in full generality, we should ideally return to studying $V$. However, since we are interested in the region where $c_H\ll 1$ we can integrate $B$ out. This returns us to the potential $V_{\Phi}$ which we arrived at above through algebraic means. By studying $V_\Phi$ instead of $V$ we miss the presence of a physical particle primarily made up of $(B-\langle B \rangle)$ whose mass is dominated by $c_B$, as well as small corrections to the masses of the other physical particles.  

Studying $V_{\Phi}$ has the advantage that practically all the hard work has been done for us; we summarize the results here (see e.g.~\cite{Haber:2015pua}). The 8 complex components in $\Phi_1$ and $\Phi_2$ get split into two CP-even real scalar singlet fields $h$ and $s$ (where we define $m_s \le m_h$), one CP-odd real scalar singlet field $A$, one charged complex scalar $H^\pm$, and three Goldstone bosons which are eaten by the gauge sector. The masses of the scalars are given by
\begin{align}m_A^2&=-\lambda_5 v^2,\\
m_{H^\pm}^2&=-\frac{1}{2}\lambda_{45} v^2,\\
\frac{m_{h,s}^2}{\mu^2}&=\frac{2\lambda}{\lambda-\lambda_{345}}(1+r)\pm \frac{2\lambda}{\lambda-\lambda_{345}}\sqrt{1 +\frac{\lambda_{345}^2}{\lambda^2}(2r+ r^2)}.
\end{align}

The CP-even Higgses $h$ and $s$ will, in general, be misaligned compared to the fields $\Phi_1$ and $\Phi_2$. As is tradition for the 2HDM, we do not actually measure the misalignment relative to $\Phi_1$ and $\Phi_2$, but relative to the fields in the Higgs basis $H_1$ and $H_2$, defined through the relation
\begin{align}
\begin{pmatrix} H_1\\ H_2\end{pmatrix}:=\begin{pmatrix}\cos\beta & \sin\beta\\ -\sin \beta & \cos \beta \end{pmatrix}\begin{pmatrix} \Phi_1\\ \Phi_2\end{pmatrix}.
\end{align}
This basis is chosen so that all the VEV sits in $H_1$, that is $\langle H_2\rangle=0$. Writing $H_i=(H_i^\pm ,H_i^0)^T$, the physical Higgses $h$ and $s$ (in the usual 2HDM parlance these are respectively called $H$ and $h$) will be a linear combination of $\sqrt{2}\mathrm{Re}H_1^0-v$ and $\sqrt{2}\mathrm{Re}H_2^0$. We define the angle $\beta-\alpha$ such that
\begin{align}
\begin{pmatrix} h\\ s\end{pmatrix}=\begin{pmatrix}\cos(\beta-\alpha) & -\sin(\beta-\alpha)\\ \sin (\beta-\alpha) & \cos (\beta-\alpha) \end{pmatrix}\begin{pmatrix}\sqrt{2} \mathrm{Re}H_1^0-v\\ \sqrt{2}\mathrm{Re}H_2^0\end{pmatrix},
\end{align}  
where we choose $(\beta-\alpha)$ such that $\cos(\beta-\alpha)\ge 0$, which fixes uniquely the definitions of $h$ and $s$. If $\cos(\beta-\alpha)=1$ the VEV and $h$ directions align, whilst if $| \sin(\beta-\alpha) | =1$ the VEV aligns with the $s$ direction. The full expression for $\sin(\beta-\alpha)$ is obtainable, however we just report it here to leading order in $r$ (around $r=0$):
\begin{align}\label{eq:sinBetaminusAlpha}
\sin(\beta-\alpha)=-\frac{\lambda-\lambda_{345}}{\lambda}\sqrt{\frac{r}{2}}+\mathcal{O}\left(r^{3/2}\right).
\end{align}
%%%%%%%%%%%%%%%%%%%%%%%
\subsection{The fermion sector} \label{sec:fermionSector}
%%%%%%%%%%%%%%%%%%%%%%%

{In the following we construct a fermion sector in which the $D_8$ symmetry is only softly broken by fermion mass terms,{ at the expense of explicitly breaking the symmetry down to $\mathbb{Z}_2$ (see the discussion at the end of Section~\ref{sec:TechnicalNatural}).} We restrict our discussion to the top sector, reducing the mention of the other fermions to the broad statement that they follow analogously.}

{In order to construct $D_8$-invariant Yukawa couplings we have to embed the RH top quark into $r_2$, i.e. we introduce a partner for the RH top quark with identical quantum numbers under the SM gauge group to obtain a full $D_8$ doublet $\tilde T_R\sim r_2$, with $\tilde T_R = (\tilde T_{1\, R}, \tilde t_R)^T$.\footnote{A tilde over a fermionic field variable marks it as a bare, i.e. non mass-eigenstate field.} Using this we can write down the $D_8$-invariant term }
\begin{align} \label{eq:D4Yuk}
\mathcal{L}_{D_8}\supset -y_t \overline{Q}_3\tilde\Phi \tilde T_R +\mathrm{h.c.}=-y_t \overline{Q}_3\tilde \Phi_1\tilde T_{1\, R}-y_t \overline{Q}_3\tilde \Phi_2\tilde t_R+\mathrm{h.c.}\,,
\end{align}
where $\tilde{\Phi}_i = i\sigma_2 \Phi^\ast$.
To raise the mass of the non-SM component we introduce a left-handed field $\tilde T_L\sim r_{++}$, which can be viewed as a vector-like partner to $\tilde T_{1\, R}$. This allows for the $D_8$-breaking but $\{1,x\}$-preserving term
\begin{align}
\mathcal{L}_{\cancel{D_8}}\supset -M_{\tilde T}\overline{\tilde T}_L \tilde T_{1\, R}+\mathrm{h.c.}
\end{align}
There are no other terms we can add in the fermionic sector which are consistent with the $\mathbb{Z}_2$ symmetry. 

After electroweak symmetry breaking, the mass terms for the top sector can be written as 
\begin{align}
\mathcal{L}\supset -\begin{pmatrix} \overline{\tilde t}_{L}& \overline{\tilde T}_L\end{pmatrix} \begin{pmatrix} y_t\frac{v}{\sqrt{2}} \sin \beta & y_t \frac{v}{\sqrt{2}}\cos\beta\\ 0 & M_{\tilde T}\end{pmatrix} \begin{pmatrix}\tilde T_{1\, R}\\ \tilde t_R\end{pmatrix}
\end{align}
ignoring a small amount of mixing between this sector and the first and second generations. After diagonalizing the mass matrix this reduces to 
\begin{align}
\mathcal{L}\supset -m_t \overline{t}_L t_R-m_T \overline{T}_L T_R
\end{align}
where $\overline{t}_L, \overline{T}_L, t_R$ and $T_R$ denote the corresponding eigenvectors, and the mass eigenvalues are given by 
\begin{align}
m_{t,T}^2=\frac{1}{2} M_{\tilde T}^2\left( 1+\frac{y_t^2 v^2}{2M_{\tilde T}^2}\mp \sqrt{1+\left(\frac{y_t^2 v^2}{2M_{\tilde T}^2}\right)^2+\frac{y_t^2 v^2}{M_{\tilde T}^2}\cos 2\beta} \right)\,,
\end{align}
which at leading order in $y_t v/ M_{\tilde T}$ reduces to
\begin{equation}
	m_t^2 = \frac{y_t^2 v^2}{2}\sin^2\beta + \mathcal{O}\bigg(\frac{y_t^2 v^2}{M_{\tilde T}^2} \bigg)\,,\qquad m_T^2 = M_{\tilde T}^2 + \mathcal{O}\bigg(\frac{y_t^2 v^2}{M_{\tilde T}^2} \bigg)\,.
\end{equation}
That is, for $y_t v/ M_{\tilde T} \ll 1$ the phenomenology of the top quark is essentially identical to the one of a type-I 2HDM.

%%%%%%%%%%%%%%%%%%%% 
\section{Phenomenology}
\label{sec:pheno}
%%%%%%%%%%%%%%%%%%%%
%

At low energies $ \ll \Lambda$ the heavy trigger field $B$ can be integrated out and the model is mapped onto a CP-conserving 2HDM, with potential $V_\Phi$ as defined above. The low-energy phenomenology of the scalar sector is therefore completely determined by the six free parameters $\{ c_H \Lambda^2, \mu^2, \lambda,\lambda_3,\lambda_4,\lambda_5\}$ of the 2HDM scalar potential. 

The only imprint of the heavy $B$ at low energies is the value of the Higgs mass parameter $c_H$ through the allowed range of $r$ values. 
We discard the region with $r \gg 1$, since that would require $\mu \ll v  \ll M_{\tilde T}$ (because $v^2 \sim \mu^2(1+r)$) instead of $\mu \sim v  \ll M_{\tilde T}$, where the second inequality is needed to ensure that the top partners evade current experimental bounds. That would again introduce a radiatively unstable large hierarchy, now between the two soft breaking parameters $\mu$ and $M_{\tilde T}$, and therefore would not provide a solution to the hierarchy problem.

When varying from $r\ll 1$ to $r\sim \mathcal{O}(1)$ one interpolates from a 2HDM which is arbitrarily close to an inert 2HDM (i2HDM), and therefore naturally in the alignment limit, to a generic 2HDM with a natural electroweak scale. For this reason we will discuss the $r\ll 1$ and $r\sim \mathcal{O}(1)$ regions separately in the following. 

Before discussing the 2HDM phenomenology in detail let us give a short overview of the main features of the two different regions of parameter space.
\begin{itemize}
	\item $\mathbf{r\ll 1:}$ This region, which we will discuss in detail in Section~\ref{eq:rll1}, is characterized by a very light CP-even Higgs with mass $m_s^2 \sim r\, m_h^2$.  
		At the same time we are automatically in the alignment limit with $|\sin(\beta-\alpha)| \ll 1$ and also $\tan\beta \gg 1$, such that couplings of $s,A,H^{\pm}$ to SM fermions and $sVV$ couplings are strongly suppressed, while couplings of the heavy CP-even mass eigenstate $h$, which we identify with the SM Higgs boson, are SM-like. Due to $\tan\beta \gg 1$ the phenomenology in this region is similar to that of an i2HDM, with the difference that $s$ is only long-lived, but not stable, and therefore does not constitute a viable dark matter candidate. The strongest constraints in this region originate from searches for invisible Higgs decays or Higgs signal strength measurements since $h\rightarrow ss,AA,H^+ H^-$ decays are unsupressed if they are kinematically accessible. Thus the prediction $m_s \ll m_h/2$ pushes us unavoidably into a slightly fine-tuned region of parameter space where $|\lambda_{345}| \ll |\lambda_{45}|$ in order to suppress $h\rightarrow ss$ decays, which are mediated by the coupling $\lambda_{345}$.
	\item $\mathbf{r\sim \mathcal{O}(1):}$ In this region the model is a generic 2HDM with a natural electroweak scale. We are not automatically pushed into the alignment limit and thus a certain amount of tuning among the parameters in the scalar potential is necessary to obtain SM-like couplings for the CP-even mass eigenstate that we identify with the SM Higgs. The SM-like Higgs can be either the lighter or heavier mass eigenstate in this scenario. {Since the phenomenology of the model in this region of parameter space is identical to the one of a completely generic 2HDM we do not discuss it in detail but refer the interested reader to Refs.~\cite{Branco:2011iw,Chowdhury:2017aav,Haber:2015pua}.}
\end{itemize}
Another phenomenologically interesting feature of the model are the vector-like fermions which are needed to construct $D_8$ invariant Yukawa couplings, as discussed in Section~\ref{sec:fermionSector}. In Section~\ref{sec:TPpheno} we estimate current bounds on their masses.
%
%%%
\subsection{Theoretical constraints and electroweak precision tests} \label{sec:TheoryEWPT}
%%%
%
Let us start the exploration of the model's phenomenology by collecting theoretical and experimental bounds which are independent of the value of $r$.

Some regions of parameter space do not lead to a theoretically consistent model. In order for the potential to be bounded from below the quartics of the scalar potential have to satisfy (see e.g.~\cite{Branco:2011iw})
\begin{equation}\label{eq:Positivity}
	\lambda > 0\,,\quad \lambda_3 > -\lambda\,,\quad \lambda_3 + \lambda_4 - |\lambda_5| > -\lambda\,.
\end{equation}
Note that these bounds are automatically satisfied for any set of parameters in the full theory for which the potential in Eq.~\eqref{eq:2HDMPotential} is bounded from below. However, in this section it is more convenient to work completely in the EFT and impose these bounds on the EFT parameters.

We also restrict the size of the quartic couplings to avoid low-scale Landau poles. This is done by solving numerically the RG equations for the $\lambda_i$ \cite{Branco:2011iw} from the electroweak scale, taken to be the $Z$ mass $M_Z$, up to some larger scale $\Lambda_{\rm Landau}$, assuming that the main contribution comes from mixing among the scalar quartics. If the sum of squared quartics exceeds $10^4$, i.e. $\sum_i \lambda_i^2 (\Lambda_{\rm Landau}) > 10^4$, which roughly corresponds to the strong coupling limit $\lambda_i \gtrsim (4\pi)^2$, we assume that at least one of the couplings has hit a Landau pole at the scale $\Lambda_{\rm Landau}$ or below and exclude the corresponding set of parameters. We have checked that the resulting exclusion contours depend only weakly on the exact value of the threshold used in the Landau pole bound. These bounds are typically stronger than tree-level perturbative unitarity bounds on the couplings.

Another set of bounds which are relevant to all regions of interest are electroweak precision tests in the form of the oblique $S,T$ and $U$ parameters~\cite{Peskin:1991sw}. The $T$ parameter especially receives considerable contributions when there is a large mass splitting between the charged and uncharged Higgses. We compute the oblique corrections following~\cite{Haber:2010bw} and cross-check our results with the publicly available code \texttt{2HDMC}~\cite{Eriksson:2009ws}. We constrain our parameter space by requiring that the contributions to $S,T$ and $U$ do not deviate by more than $2\sigma$ from the PDG values~\cite{ParticleDataGroup:2020ssz}
\begin{equation}
	\hat{S}= -0.01\pm 0.10\,,\quad \hat{T}=0.03\pm 0.12\,,\quad \hat{U}=0.02\pm 0.11\,,\quad \rho=\begin{pmatrix}
		1	&	0.92	&	-0.80\\
		0.92	&	1	&	-0.93\\
		-0.80	&	-0.93	&	1
	\end{pmatrix}\,,
\end{equation}
where $\rho$ is the correlation matrix. This is achieved by evaluating
\begin{equation}
	\chi^2_{STU} = \mathbf{x}^T V^{-1}\mathbf{x}\,,
\end{equation}
with the covariance matrix $V$ corresponding to $\rho$ and $\mathbf{x}=(S-\hat{S},T-\hat{T},U-\hat{U})^T$, and demanding that $\chi^2_{STU} \leq 8.03$, corresponding to deviation of at most $2\sigma$. 

Note that electroweak fits which include a recent measurement of the $W$ mass by the CDF collaboration at the Tevatron~\cite{CDF:2022hxs} prefer larger values of the $S$ and $T$ parameter (see e.g.~\cite{Asadi:2022xiy}). However, since the reported value is considerably higher than in previous measurements at the Tevatron and LEP~\cite{CDF:2013dpa}, ATLAS~\cite{ATLAS:2017rzl} and LHCb~\cite{LHCb:2021bjt} and is in serious tension with the SM prediction, we take a conservative approach and compute bounds based on the PDG values as outlined above. %However, should the $W$ mass measurement by the CDF collaboration be confirmed our model can easily accommodate 
%
%%%
\subsection{$r \ll 1$: alignment region}\label{eq:rll1}
%%%
%

In the limit $r\rightarrow 0$ we arrive at an i2HDM which is characterized by SM-like Higgs couplings and an unbroken $\mathbb{Z}_2$ parity under which all BSM Higgses, i.e. $\{s,H^\pm,A\}$, and fermion partners are odd and all SM particles even. Thus all couplings with an odd number of BSM Higgses are suppressed by at least $v_1 / v \sim\sqrt{r}$ (cf. Eq.~\eqref{eq:tanBeta}). In addition the mass of the lighter CP-even scalar $m_s$ is much less than the one of the heavier, SM-like Higgs; in particular $m_s^2/m_h^2\sim r$.  As such the phenomenology splits into two essentially disconnected parts, the first focusing on $s$ --- for which the heavy scalars are irrelevant --- and the second on the CP-odd and charged Higgses $A$ and $H^\pm$. In the study of $A$ and $H^\pm$ the light Higgs scalar can effectively be taken as massless.

\paragraph{The phenomenology related to the light Higgs scalar $s$: } {There are broadly two classes of experimental bounds on $s$. The first class of experimental probes is related to unsurpressed trilinear Higgs couplings containing $s$, and their effect on Higgs precision data.}
The second consists of statements relating to the interaction of $s$ with fermions, which can be experimentally probed in flavor precision measurements. Let us start by exploring the trilinear couplings.

Neglecting terms suppressed by $r$, the only trilinear coupling of $s$ to the other Higgses is
\begin{align}\label{eq:ssh}
-\frac{1}{2} \lambda_{345} v s^2 h.
\end{align}
This puts strong constraints on $\lambda_{345}$ since the decay $h\rightarrow ss$ is always kinematically accessible in the $r\ll 1$ region. Depending on the lifetime of $s$ it either decays within the detector and modifies the Higgs signal strength or escapes the detector and contributes to the invisible Higgs width. Current global Higgs signal strength measurements are  $ 1.06 \pm 0.07$ at ATLAS~\cite{ATLAS:2020qdt} and $ 1.02^{+0.07}_{-0.06}$ at CMS~\cite{CMS:2020gsy}, whereas the invisible Higgs width is constrained to be BR$(h\rightarrow \text{inv}) < 0.15$ at $95\%$ CL at ATLAS~\cite{ATLAS:2022yvh} and BR$(h\rightarrow \text{inv}) < 0.18$ at CMS~\cite{CMS:2022qva}. All of these give approximately the same bound
\begin{equation}\label{eq:lambda345Small}
	|\lambda_{345}| \lesssim 0.01\,.
\end{equation}
Thus in general $|\lambda_{345}|\ll |\lambda_{45}|$, which corresponds to a tuning since $\lambda_{345}=0$ does not lead to an enhanced symmetry. Conversely, we do not have to tune the parameters to reach the alignment limit, which is automatic in this region of parameter space. We will comment more on the tuning in Section~\ref{sec:Tuning}.

In order to determine the experimental signals that $s$ will give, it is important to keep in mind that its linear couplings to the SM are suppressed by powers of $m_s/m_h$. This suppression is strong enough such that $s$ is long-lived in a large region of parameter space. The phenomenologically relevant couplings are of the form
\begin{equation}
	\mathcal{L}_{s} \supset  C_{sVV} s\left( \frac{2 m_W^2}{v} W_\mu^- W^{+\, \mu}+\frac{m_Z^2}{v} Z_\mu^2\right) - C_{sff} \frac{m_f}{v}  s \bar{f} f + C_{sH^\pm} \frac{2 m_{H^\pm}^2}{v} s H^+ H^-\,,
\end{equation}
with
\begin{equation}
	C_{sVV} = \sin(\beta -\alpha)\approx 
	-  \frac{m_s}{m_h}
	\,,\quad \quad C_{sff} = \frac{\cos \alpha}{\sin\beta} \approx
	\frac{\lambda_{345}}{\lambda}\frac{m_s}{m_h}\,,\quad C_{sH^\pm } \approx \left( 1 - \frac{\lambda}{\lambda_{45}}\right)\frac{m_s}{m_h}\,,
\end{equation}
where we assumed that $|\lambda_{345}|\ll \lambda$ and omitted the previously discussed $ssh$ coupling in Eq.~\eqref{eq:ssh} and the coupling to the top partners. At energies $E\sim m_s \ll v$ it is convenient to work with an effective Lagrangian in which the top quark, its partner, the W- and Z-boson and the heavy Higgses are integrated out~\cite{Kniehl:1995tn}
\begin{equation}\label{eq:LagsEff}
	\mathcal{L}_{\rm eff}= - \frac{m_f}{v} C_{sff} s \bar{f} f  + \frac{C_{sgg}\, \alpha_s}{12\pi}\frac{s}{v} G^a_{\mu\nu} G^{a\, \mu\nu} + \frac{C_{s\gamma\gamma}\, \alpha_{\rm em}}{2\pi}\frac{s}{v}F_{\mu\nu}F^{\mu\nu}\,,
\end{equation}
with
\begin{equation}
	C_{sgg} = C_{sff} \,,\qquad C_{s\gamma\gamma} =Q_t^2 C_{stt} -\frac{7}{4} C_{sVV} - \frac{1}{12}C_{sH^\pm}\,,
\end{equation}
where $Q_t = 2/3$ is the charge of the top quark. The Wilson coefficients $C_{sgg}$ and $C_{s\gamma\gamma}$ receive one-loop contributions when integrating out the top and its partner as well as the electrically charged Higgs and W-boson.\footnote{Note that $C_{sgg}$ and $C_{s\gamma\gamma}$ are insensitive to parameters of the top partner. This is a well-known phenomenon in the Composite Higgs literature and can be traced back to the fact that our model allows for only one invariant that generates the top mass~\cite{Azatov:2011qy,Montull:2013mla}.} Note that Eq.~\eqref{eq:LagsEff} is essentially the low-energy Lagrangian of a real scalar mixing with the Higgs through a small mixing angle $C_{sff} \sim m_s/m_h \ll 1$. This is similar to the phenomenology of the crunching dilaton model~\cite{Csaki:2020zqz}, which features a light scalar, the dilaton, weakly mixing with the Higgs. The principal difference is that the dilaton of Ref.~\cite{Csaki:2020zqz} has an additional tree-level coupling to the photon.

\begin{figure}[t]
\begin{center}
\includegraphics[width=0.48\textwidth]{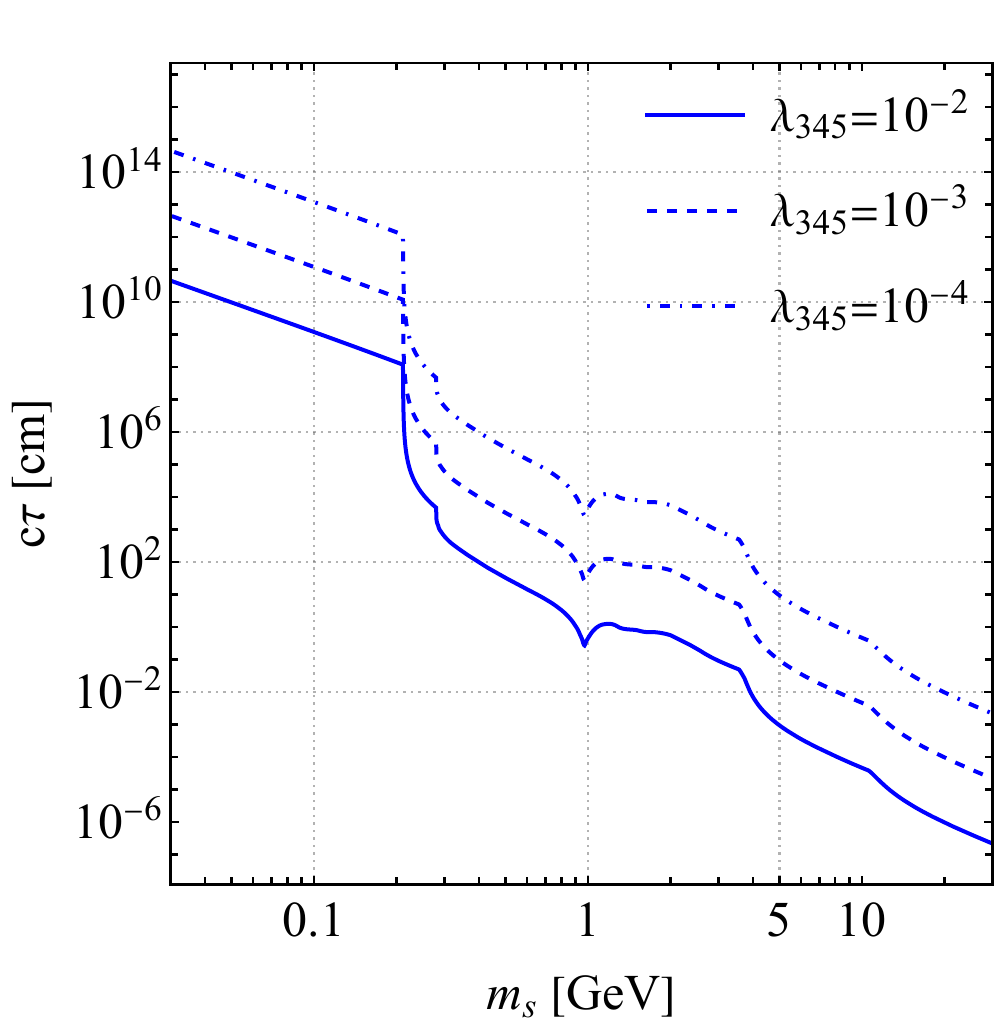}
\includegraphics[width=0.48\textwidth]{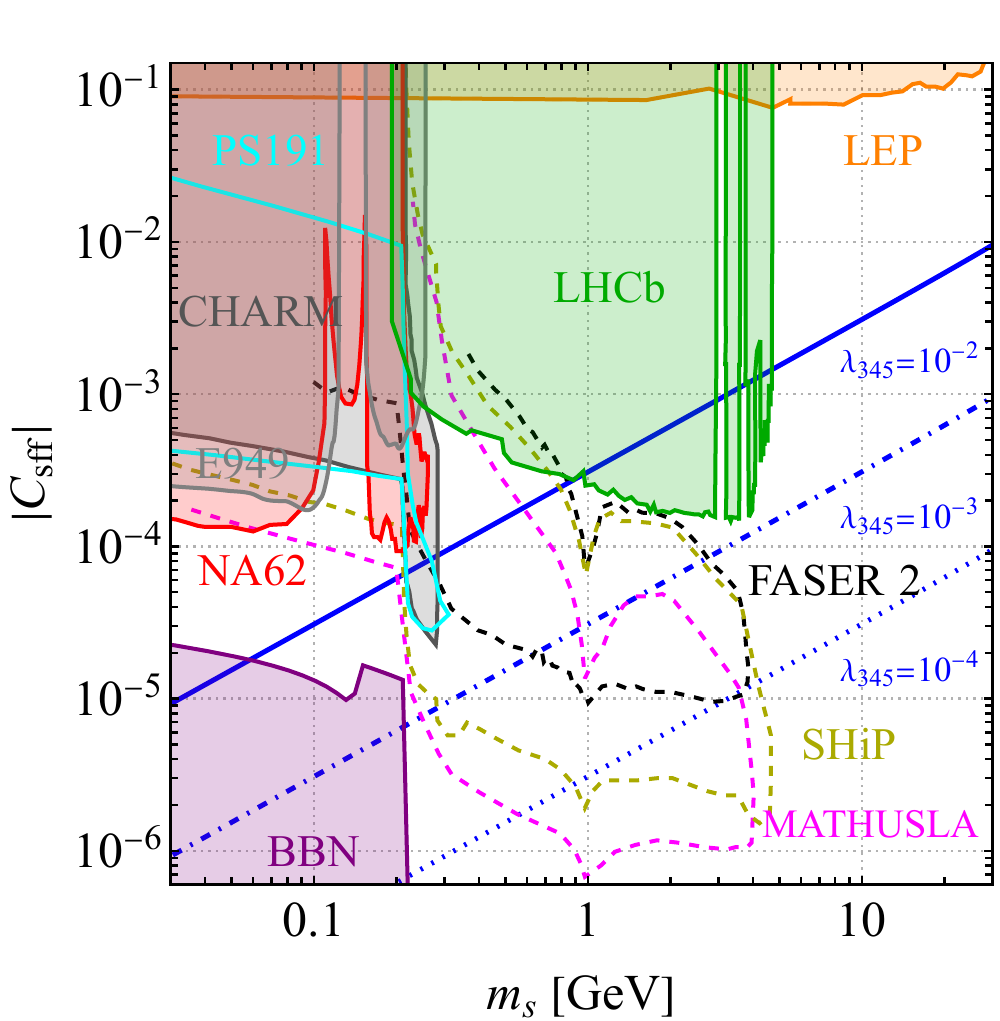}
\end{center}
\caption{Left: proper lifetime contour along  $10^{-7} \leq r \leq 10^{-1}$ of the light CP-even Higgs $s$ for various values of $\lambda_{345}$. The decay widths into hadronic final states are taken from~\cite{Winkler:2018qyg}. Right: bounds on the light CP-even Higgs in the $|C_{sff}|$-$m_h$ plane. The blue lines show the same contours as in the left panel, where we vary $10^{-7} \leq r \leq 10^{-1}$ and fix $\lambda$ and $\mu^2$ such that we reproduce $m_h = 125$ GeV and $v=246$ GeV. For an overview of the bounds and their origin see the main text.}
\label{fig:lowHiggsMass}
\end{figure}

Using this Lagrangian we can determine the lifetime of $s$, which we show as a function of its mass $m_s$ in the left panel of Figure~\ref{fig:lowHiggsMass}. Hadronic final state contributions for $m_{s}\lesssim 2$ GeV are taken from~\cite{Winkler:2018qyg}. The left panel of Figure~\ref{fig:lowHiggsMass} shows that the light Higgs is indeed long-lived over a large mass range. This is due to the $m_s / m_h \ll 1$ suppression of the couplings to the SM and the small Yukawa couplings to the kinematically accessible final states. Note that decays to photons only give a subleading contribution to the decay width over the entire mass range shown in the plot.

With the lifetime at hand, we can now study experimental probes of the light Higgs. These are mainly sensitive to the coupling to SM fermions. Thus, it is pertinent to study the $C_{sff}$ vs. $m_s$ parameter space. At low masses ($m_s \lesssim 5$ GeV) the light Higgs affects meson decays and thus the strongest constraints come from flavor precision measurements. Also note that $s$ is long-lived in the lower mass region (see left panel of Figure~\ref{fig:lowHiggsMass}) and therefore it causes displaced decay vertices or escapes the detector without decaying and shows up as missing energy. 

In the right panel of Figure~\ref{fig:lowHiggsMass} we collect all bounds on the scalar-fermion coupling. However, note that we only show the tightest constraints with subleading constraints being available in~\cite{Winkler:2018qyg,Goudzovski:2022vbt}. 
The blue lines show the model prediction for the scalar-fermion coupling $|C_{sff}|$ with $|\lambda_{345}| = 10^{-2},10^{-3},10^{-4}$ as a function of the mass $m_s$ for $r$ in the range $10^{-7} \leq r \leq 10^{-1}$ and $\lambda$, $\mu^2$ fixed such that we reproduce $m_h = 125$ GeV and $v=246$ GeV. We now give a detailed account of all the individual bounds that entered the figure:

\begin{itemize}

\item {\bf LEP:} LEP searches for neutral Higgs bosons in the $e^+ e^-\rightarrow Z^* h$ channel with hadronic $Z$ decays \emph{(orange band)}~\cite{L3:1996ome} are sensitive to light scalars. At masses $m_s \gtrsim 5$ GeV they are the strongest constraint on our model.

\item {\bf B decays:} The strongest bounds from $B$ decays originate from searches for displaced $B \rightarrow h \mu^+ \mu^-$ decays at LHCb \emph{(green band)}~\cite{LHCb:2015nkv,LHCb:2016awg}. The bound shown in Figure~\ref{fig:lowHiggsMass} is adapted from~\cite{Winkler:2018qyg}. 

\item {\bf Kaon decays:} Below the muon threshold $m_s < 2 m_\mu$ searches for rare Kaon decays are most sensitive to our scenario. Note that in this region because of its long lifetime $s$ usually escapes the detector before decaying and shows up as missing energy. The bounds originate from $K\rightarrow \pi + X$ searches with invisible $X$ measured by the NA62 collaboration \emph{(red band)}~\cite{NA62:2020pwi,NA62:2020xlg,NA62:2021zjw} and the E949 collaboration \emph{(silver band)}~\cite{BNL-E949:2009dza}. We also show recasts of the CHARM beam dump experiment \emph{(gray band)}~\cite{CHARM:1985anb} adapted from~\cite{Goudzovski:2022vbt} and of PS191 \emph{(cyan band)}~\cite{Gorbunov:2021ccu}.

\item {\bf Astrophysical bounds:} There are also astrophysical and cosmological constraints. The extraordinary success of BBN in predicting the abundances of light elements constrains the lifetime of the light scalar. In order not to spoil the well-established predictions, $s$ decays cannot inject a considerable amount of energy during BBN. Requiring that the $s$ abundance decays before BBN occurs puts a bound on its lifetime of roughly $\tau \lesssim 1$~s. The exact BBN bound \emph{(purple band)} we show depends on the decay channels and is taken from~\cite{Fradette:2017sdd}. There are also bounds from supernova cooling~\cite{Krnjaic:2015mbs,Evans:2017kti,Dev:2020eam} which are subleading to BBN bounds in the region of parameter space we are interested in.

\item {\bf Future sensitivity:} FASER 2 \emph{(black line)}~\cite{Anchordoqui:2021ghd}, MATHUSLA \emph{(yellow line)}~\cite{MATHUSLA:2020uve,MATHUSLA:2022sze} and SHiP \emph{(magenta line)}~\cite{SHiP:2015vad} will be able to probe a large region of parameter space for $m_s > 2m_\mu$. However, there is no current or planned experiment which will close the gap between NA62 and the BBN bound. This would require to improve the sensitivity on $BR(K^+ \rightarrow \pi^+ h)$ from $BR(K^+ \rightarrow \pi^+ h)\approx 10^{-11}$ currently reached by the NA62 collaboration for $m_h < 2m_\mu$ down to $BR(K^+ \rightarrow \pi^+ h)\approx 10^{-13}$~\cite{Goudzovski:2022vbt}.

\end{itemize}

\begin{figure}[t]
\centering
\includegraphics[width=0.6\textwidth]{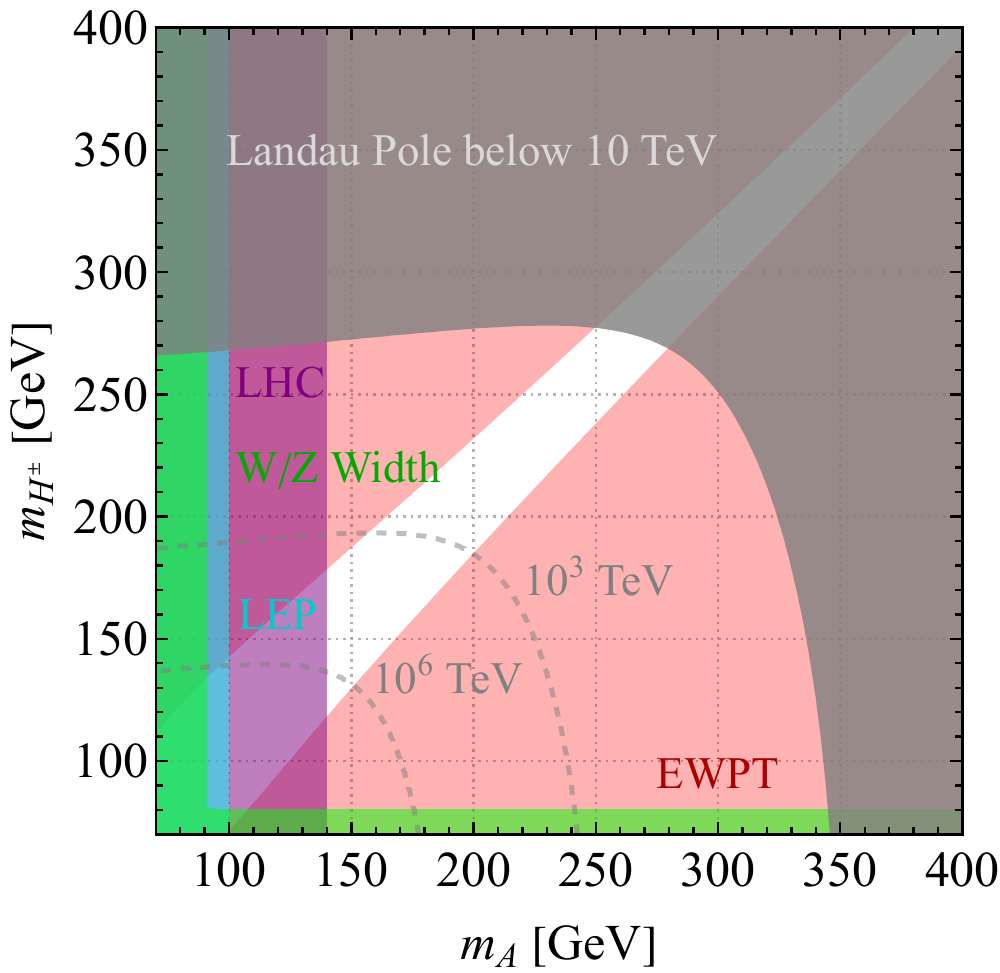}
\caption{Experimental constraints on the charged and pseudoscalar Higgs masses in the $r\ll 1$ regime for $m_h, \sin(\beta-\alpha), \tan^{-1}\beta\approx 0$ and $\lambda_{345}=0.01$ fixed to its maximal allowed value. The white region shows the allowed parameter space. In red we show the $2\sigma$ bounds from electroweak precision tests as discussed in Section~\ref{sec:TheoryEWPT}. The region shaded in gray corresponds to parameter values which develop a Landau pole in the scalar quartics below $10$ TeV. For reference we also show as dashed lines how far this region would extend in order to prevent a Landau pole below a scale of $10^3,10^6$ TeV. The green region is excluded by $W$ and $Z$ width measurements which forbid unsuppressed $W^\pm\rightarrow H^\pm h$ and $Z\rightarrow h A$ decays. The cyan and purple shaded regions are bounds from recast LEP-II limits for neutralino production~\cite{Lundstrom:2008ai} and LHC dilepton searches~\cite{Belanger:2015kga}, respectively.}
\label{fig:lowHiggsMassheavyHiggs}
\end{figure}

\paragraph{The phenomenology related to $H^\pm$ and $A$: }Let us now turn to bounds on the charged and pseudoscalar Higgses $H^\pm$ and $A$. In the $r\ll 1$ regime their phenomenology is to a good approximation completely determined by $\{\lambda_{45},\lambda_5\}$ or equivalently their masses $\{m_{H^\pm},m_A\}$. This is the case since we can effectively take $m_s/m_h,\sin(\beta-\alpha),\tan^{-1}\beta \approx 0$, and $m_h,v$ fixed to their SM values. Under these assumptions 
one retains an approximate $\mathbb{Z}_2$ symmetry (which is exact when $r=0$) under which $\Phi_1$ and the fermion partners are odd. This symmetry 
 strongly suppresses couplings with an odd number of BSM Higgses $\{s,H^\pm,A\}$ and makes the phenomenology for $A$ and $H^\pm$ practically identical to that of an i2HDM. Thus, in the following we will assume that $s$ is stable and invisible in collider searches. We collect all constraints in the $m_{H^\pm}$-$m_A$ plane in Figure~\ref{fig:lowHiggsMassheavyHiggs}.
\begin{itemize} 
    \item {\bf EWPT:} We show constraints from electroweak precision observables \emph{(red bands)} and from requiring that there is no Landau pole in the scalar quartics below $10$ TeV \emph{(gray band, $10^3,10^6$ \emph{TeV as} dashed gray lines)}, following the strategy we explained in Section~\ref{sec:TheoryEWPT}. The strong bounds from EWPT are mainly driven by large contributions to the $T$ parameter which prefers $m_A \approx m_{H^\pm}$.

\item {\bf W/Z width:}  \sloppy  Strong constraints also originate from the unsuppressed $W^\pm H^\mp s, W^\pm H^\mp A, Z s A,Z H^+ H^-$ couplings which modify the well-measured $W$ and $Z$ decay widths \cite{ALEPH:2005ab,Janot:2019oyi,ALEPH:2013dgf,TevatronElectroweakWorkingGroup:2010mao} if these decays are kinematically accessible. In order to avoid these constraints one has to require
\begin{equation}
	m_h + m_{H^\pm} > m_W\,,\quad m_A + m_{H^\pm} > m_W\,, \quad m_h + m_A > m_Z\,,\quad 2 m_{H^\pm} > m_Z\,,
\end{equation}
        which for the current setup approximately translates into $m_A > m_Z$ and $m_{H^\pm} > m_W$ and is shown in Figure~\ref{fig:lowHiggsMassheavyHiggs} \emph{(green band)}.

\item {\bf LEP:} Ref.~\cite{Lundstrom:2008ai} performed a reinterpretation of LEP-II limits for neutralino production in terms of the i2HDM and found a limit of
\begin{equation}\label{eq:mA_bound_cyan}
	m_A > 100\text{ GeV}
\end{equation}
for $m_A - m_s > 8$~GeV, which we show in Figure~\ref{fig:lowHiggsMassheavyHiggs} \emph{(cyan band)}.

LEP-II is also sensitive to charged Higgs pair production $e^+ e^-\rightarrow H^+ H^-$ which results in a bound of
\begin{equation}
	m_{H^\pm} > 70\text{ GeV}\,,
\end{equation}
which was found in a recast of LEP bounds on charginos in~\cite{Pierce:2007ut}. This constraint does not show up on our plot, however.

\item {\bf LHC:} An even stronger bound on $m_A$ than Eq.~\eqref{eq:mA_bound_cyan}, which we show in Figure~\ref{fig:lowHiggsMassheavyHiggs} \emph{(purple band)}, is obtained from a recast of dilepton searches at LHC run 1~\cite{Belanger:2015kga} which results in
\begin{equation}
	m_A > 130-140\text{ GeV}
\end{equation}
for $m_s \approx 0$, where the exact bound weakly depends on $m_{H^\pm}$. To our knowledge there is no updated analysis with run 2 data. A simple rescaling of the bound shows that it could improve up to $m_A > 170$ GeV for approximately massless $s$. However, in order to set a robust bound a dedicated analysis is needed.
\end{itemize}
As can be seen from Figure~\ref{fig:lowHiggsMassheavyHiggs} the allowed parameter space in the $r\ll 1$ regime comprises  exotic Higgses in the mass range $130\text{ GeV} \lesssim m_A, m_{H^\pm} \lesssim 270$ GeV with a small mass splitting. 
%
%
%%%

%%%
\subsection{Fermion partners}\label{sec:TPpheno}
%%%
%
As discussed in Section~\ref{sec:fermionSector}, to have a consistent theory with small $\mu$, the fermionic sector requires partner fermions. The top partners are phenomenologically most relevant, and what we shall discuss here. 

The prominent top partner decay channels are through $W^+b, Zt, ht, H^+ t, A t, st$. We first estimate the branching ratios of these decays and then use them to find experimental constraints on $m_T$ from top partner pair production. 

\paragraph{Top partner decay channels:} When the mass of the top partner $m_T$ is much greater than the $W$ mass, the Goldstone boson equivalence theorem allows its branching ratios to be found from its couplings to the Higgs scalars and the SM top. To leading order in the (assumed small) quantity $m_t/m_T$ these couplings are given by
\begin{equation}\label{eq:FermionLagrangian}
\begin{split}
	\mathcal{L} \supset &\frac{\sqrt{2} m_t}{v}\bar{b}_L\left[(G^- + \cot \beta \, H^-) t_R - (H^- - \cot\beta\, G^-)T_R \right]  - m_T \bar{T}_L T_R\\
        &-\frac{m_t}{v} \bar{t}_L T_R \left(\frac{\cos\alpha}{\sin\beta} h -\frac{\sin\alpha}{\sin\beta} s +\mathrm{i} A -\mathrm{i}\cot \beta\,  G^0 \right) + \mathrm{h.c.}
	\end{split}
\end{equation}
where $G^\pm$ and $G^0$ are the Goldstone bosons which are eaten by $W^\pm$ and $Z$, respectively. 
Working close to the alignment limit\footnote{Note that $\cos(\beta - \alpha)\approx 1$ corresponds to the alignment limit when the heavier CP-even mass eigenstate is identified with the SM-like Higgs boson.} $\cos(\beta - \alpha)\approx 1$, the corresponding decay branching ratios of $T$ are approximately given by
\begin{align}
&\text{BR}(T \rightarrow W^+ b)=2\,\text{BR}(T\rightarrow Z\, t)= 2\, \text{BR}(T\rightarrow h\, t)= \frac{1}{2+2\tan^2\beta}\,,\\
	&\text{BR}(T \rightarrow H^+ b)=2\,\text{BR}(T\rightarrow A\, t)= 2\, \text{BR}(T\rightarrow s\, t)= \frac{\tan^2\beta}{2+2\tan^2\beta}\,,
\end{align}
where the first line are the standard decay channels of a $SU(2)_L$ singlet vector-like quark and the second line collects all decays into exotic Higgses.

The main production channel for the top partner at hadron colliders is through QCD pair production, thus we focus on bounds on the top partners arising from this.\footnote{Note that the cross-section for single production via vector boson exchange is always suppressed by $\cot^2\beta$.} 
Further, since the decay signature of the top partners strongly depends on $\tan\beta$ and the properties of the light Higgs $s$ we discuss the  $r\ll 1$ and $r\sim \mathcal{O}(1)$ region separately.

\paragraph{$\mathbf{r\ll 1}:$} Since in this region $\tan \beta \gg 1$, $T$ almost exclusively decays into $H^+b$, $At$, or $st$. This can be understood with the help of the approximate $\mathbb{Z}_2$ symmetry of the Higgs sector for $r\ll 1$ under which the BSM Higgses are odd. The $\mathbb{Z}_2$ is also a symmetry of the Yukawa couplings in Eq.~\eqref{eq:D4Yuk} under which $T$ is odd. Thus only decays of $T$ into BSM Higgses are unsupressed and the scalar $s$, as the lightest $\mathbb{Z}_2$-odd particle, will be at the end of the decay chain.

In addition $s$ is typically long-lived, which will result in displaced decay vertices or the light Higgs escaping the detector completely. In the following we will assume that $s$ escapes the detector and shows up as missing energy. This is an excellent approximation for $m_s\lesssim 0.3$ GeV. In this scenario the signature strongly resembles that of stop pair production $pp\rightarrow \tilde{t}_1 \tilde{t}_1^*$ with subsequent decay either directly into the lightest neutralino $\tilde{t}_1 \rightarrow t \tilde{\chi}_1^0$ or via charginos $\tilde{t}_1 \rightarrow b \tilde{\chi}_1^+ \rightarrow b W^+ \tilde{\chi}_1^0$, where $s$ plays the role of the neutralino. As the topology of these processes is similar to $pp\rightarrow \bar{T} T$ with either $T\rightarrow t s$ or $T\rightarrow b H^+  \rightarrow b W^+ s$, we perform a crude estimate for the bound on the top partner masses by taking into account the different production cross-sections for colored fermions and scalars as well as the branching ratios in our model. However, note that this is only an order of magnitude estimate as the shapes of kinematic distributions are affected by the top partners being fermions rather than scalars; a dedicated analysis would be required to obtain a robust bound. The currently strongest bounds on stop pair production are set by a combination of CMS searches at a center of mass energy of $13$ TeV and integrated luminosity of $137$ fb$^{-1}$~\cite{CMS:2021eha}.

In the $pp\rightarrow (\tilde{t}_1 \rightarrow t \tilde{\chi}_1^0)(\tilde{t}_1^* \rightarrow \bar{t} \tilde{\chi}_1^0{}^*)$ channel with a massless neutralino a bound of $m_{\tilde{t}_1} > 1325$ GeV at $95\%$ CL is quoted, assuming branching ratios of $1$. We can convert this into a bound on the top partner mass $m_T$ by equating the stop production cross-section associated to the stop mass $	\sigma_{pp\rightarrow \tilde{t}_1^*\tilde{t}_1} (1325\text{ GeV})$ with the cross-section for top partner production in the corresponding channel, i.e. we have to solve
\begin{align}
\text{BR}(T\rightarrow ts)^2\, \sigma_{pp\rightarrow \bar{T} T} (m_T) = \sigma_{pp\rightarrow \tilde{t}_1^*\tilde{t}_1} (1325\text{ GeV})\,.
\end{align}
We compute $\sigma_{pp\rightarrow \bar{T} T} (m_T)$ using HATHOR~\cite{Aliev:2010zk} and extract the stop production cross-section from~\cite{Borschensky:2014cia}. Solving the above equation for $m_T$ yields the bound
\begin{align} \label{eq:TPbound1}
m_T> 1310~\mathrm{GeV}.
\end{align}

The $pp\rightarrow (\tilde{t}_1 \rightarrow b W^+ \tilde{\chi}_1^0)(\tilde{t}_1^* \rightarrow \bar{b} W^- \tilde{\chi}_1^0{}^*)$ channel, on the other hand, yields a bound of $m_{\tilde{t}_1} > 1260$ GeV at $95\%$ CL on the stop mass, again assuming a massless neutralino and branching ratios of $1$. Analogously to the first channel we can convert this to a bound on $m_T$ by solving
\begin{align}
\text{BR}(T\rightarrow b H^+)^2\,\text{BR}(H^+\rightarrow W^+ s)^2\,\sigma_{pp\rightarrow \bar{T} T} (m_T)= \sigma_{pp\rightarrow \tilde{t}_1^*\tilde{t}_1} (1260\text{ GeV})\,.
\end{align}
On the assumption that $\text{BR}(H^+\rightarrow W^+ s)\approx 1$, this gives a lower bound of 
\begin{align}
m_T> 1360~\mathrm{GeV}\,,
\end{align}
which is slightly stronger than the bound in Eq.~\eqref{eq:TPbound1}.

\paragraph{$\mathbf{r\sim \mathcal{O}(1)}:$} In this regime the light Higgs is generically heavier $m_s \gtrsim 1$ GeV and no longer long-lived on detector timescales. Additionally the $\cot\beta$ suppression is less severe and some of the typical top partner decay channels, such as $T\rightarrow t h$, $T\rightarrow b W^+, T\rightarrow t Z$, open up. However, for $\tan\beta \geq 1$ the fraction of these ``typical'' top partner decays compared to all decays is at most $50\%$ and generally considerably smaller. The currently strongest bounds on $SU(2)_L$ singlet top partners are set by a combined analysis performed by the ATLAS collaboration on $36.1$ fb$^{-1}$ of data collected at a center of mass energy of $13$ TeV~\cite{ATLAS:2018ziw}. This analysis finds a limit of $m_T > 1300$ GeV at $95\%$ CL if one assumes that $\mathrm{BR}(T\rightarrow t h,b W^+,t Z)=1$. Adjusting this constraint by a na\"ive rescaling of the top partner production cross-section by $\mathrm{BR}(T\rightarrow t h,b W^+,t Z)^2\leq 1/4$, this yields a bound of $m_T \gtrsim 1$ TeV for the largest possible branching ratio, i.e. $\mathrm{BR}(T\rightarrow t h,b W^+,t Z)= 1/2$. A more accurate bound would require a dedicated study which takes into account the decays into exotic Higgs states and is beyond the scope of the present work.

%
%%%
\subsection{Tuning}\label{sec:Tuning}
%%%
%
Even though our model is able to explain a large hierarchy between the electroweak scale and the cutoff of the theory, some residual tuning is nevertheless required to comply with experimental observations. This tuning has two major sources which independently arise in the scalar and in the fermion sector. While the tuning in the scalar sector mainly originates from the requirement of having a SM-like  Higgs boson, the tuning in the fermion sector is due to the little hierarchy $M_{\tilde T} \gg |\mu| \sim m_H$, which is forced on us by the nonobservation of top partners at the LHC. The top partners are required to make the Yukawa couplings $D_8$ invariant, i.e. the top partners cancel the quadratically divergent contribution of the top quark to $\mu^2$. Note that we share this little hierarchy with other solutions of the hierarchy problem which have colored fermion partners in their spectrum. Let us also stress that we do not require gauge partners as the gauge couplings automatically respect the $D_8$ symmetry by construction. In the following we will give a qualitative overview of the required tuning.
\paragraph{Tuning in the scalar sector:}
Depending on the region of parameter space, the tuning in the scalar sector is typically dominated by either reaching the vicinity of the alignment limit or by suppressing Higgs decays into the light scalar $s$. In the following we give a short overview of both sources of tuning.
\begin{itemize}

\item {\bf Reaching the alignment limit:} We have two experimental indications that our 2HDM should be close to the alignment limit, in which $\sin (\beta - \alpha )=0$ and the heavy CP-even Higgs couplings are SM-like. The first is current LHC measurements of the Higgs couplings which can deviate from their SM values by roughly $10\%$~\cite{CMS:2020gsy,ATLAS:2020qdt}, indicating that $|\sin (\beta - \alpha )| \lesssim 0.1$. The second is fits within the 2HDM which prefer a smaller deviation from the alignment limit corresponding to $|\sin (\beta - \alpha )| \lesssim 0.03$~\cite{Chowdhury:2017aav}. Thus, we must be near the alignment limit.

The vicinity of the alignment limit is reached in different ways in the $r\ll 1$ and $r\sim \mathcal{O}(1)$ regions. In the $r\ll 1$ region we automatically have $|\sin(\beta - \alpha )|\ll1$. However, some amount of tuning is required when $r\sim \mathcal{O}(1)$ as this region of parameter space corresponds to a generic 2HDM. The amount of tuning which is necessary in such a scenario has been studied in detail in~\cite{Bernal:2022wct}. They find that the tuning scales as $(|\sin (\beta -\alpha)|\, \tan\beta)^{-1}$ 
and thus to get $|\sin (\beta -\alpha)|\sim 0.01$--$0.1$ we need a tuning of approximately $1\%$--$10\%$.\footnote{In particular, they use the Barbieri-Giudice definition of tuning. This quantifies the logarithmic variation of a quantity $\Omega$ with respect to its input parameters $\theta_i$. The tuning is given by $\Delta\Omega:= \max\left|\frac{d\log \Omega}{d\log \theta_i}\right|$.}

\item {\bf Suppressing Higgs decays:} There is an additional source of tuning in the scalar sector when $m_s < m_h / 2$ since the decay of the SM-like Higgs into the light CP-even mass eigenstate, $h \rightarrow ss$, is kinematically allowed. This decay is mediated by the $\lambda_{345}$ coupling and is therefore not suppressed by a small mixing angle, $\sin(\beta-\alpha)$, or by $1/\tan\beta$. In the $r\ll 1$ regime, where the decay is always kinematically allowed, to avoid large contributions to the Higgs signal strength or to the Higgs invisible width we  require that $|\lambda_{345}|\ll |\lambda_{45}|$. Due to the lower bound on the mass of $H^\pm$ (see Section~\ref{eq:rll1}) the minimal amount of tuning is given by $\frac{|\lambda_{345}|}{|\lambda_{45}|} \lesssim 1.8\%$. Note that this source of tuning disappears when $m_s > m_h/2$ which is typically the case for $r\sim\mathcal{O}(1)$.

\end{itemize}
In summary the tuning in the scalar sector is of the order of $1\%$--$10\%$ for  both regions of $r$. For $r\ll 1$ the tuning is dominated by the requirement that $|\lambda_{345}|\ll |\lambda_{45}|$ to suppress $h\rightarrow ss$ decays, whereas for $r\sim \mathcal{O}(1)$ the tuning is needed to reach the vicinity of the alignment limit.
\paragraph{Tuning in the fermion sector:}
We now estimate the amount of tuning that is required to push the top partner mass up to the TeV scale (see Section~\ref{sec:TPpheno}). Due to the explicit breaking of the $D_8$ symmetry by the bare top partner mass $M_{\tilde T}$, one-loop corrections to the mass coefficients of $\Phi_1^\dagger \Phi_1$ and $\Phi_2^\dagger \Phi_2$ are not symmetric and thus give a logarithmically divergent contribution to $\mu^2$. The combined effect of top and top partner loops give a contribution of
\begin{equation}\label{eq:muRen}
	\delta\mu^2 = \frac{N_c y_t^2 M_{\tilde T}^2}{8\pi^2}\log\frac{\Lambda}{M_{\tilde T}}\,,
\end{equation}
where $\Lambda$ is the cutoff of the theory. The other fermionic partners contribute to $\delta \mu^2$, however since the contribution only depends on the product of Yukawa coupling and fermion partner mass, e.g. $y_t M_{\tilde T}$, the fermionic partners for lighter quarks or leptons with $y_f \ll y_t$ can be naturally heavier without introducing additional tuning. For this reason we assume that the tuning in the fermion sector is dominated by the leading top partner contribution.

In both the $r\ll1$ and $r\sim \mathcal{O}(1)$ regions, $\mu$ is generically of the order of the SM-like Higgs mass $m_h$. In fact, up to small $ \mathcal{O}(\lambda_{345})$ corrections in the $r\ll 1$ region we find that $\mu = m_h /2 =62.5$ GeV. In the $r\sim\mathcal{O}(1)$ region $\mu$ and $m_h/2$ can be separated by an $\mathcal{O}(1)$ factor. The important point for the tuning is that $\mu\sim m_h$ is significantly smaller than the radiative corrections in Eq.~\eqref{eq:muRen} from TeV-scale top partners.  Thus we require a cancellation with a tree-level contribution such that $\mu^2 = \mu_{\rm tree}^2 + \delta \mu^2$. We quantify the required tuning between the contributions as
\begin{equation}
	\frac{\mu^2}{\delta \mu^2} = 3.6\cdot 10^{-2} \cdot \bigg(\frac{\mu}{125\text{ GeV}}\bigg)^2 \bigg(\frac{1\text{ TeV}}{M_{\tilde T}}\bigg)^2 \bigg(\frac{5}{x}\bigg)\,,
\end{equation}
where $\Lambda / M_{\tilde T} = 10^x$. TeV-scale top partners therefore require around a percent-level tuning.
%

%%%%%%%%%%%%%%%%%%%% 
\section{Neutrino masses} \label{sec:Neutrinos}
%%%%%%%%%%%%%%%%%%%%
A key  feature of our model is the presence of the SM-singlet field $B$ which obtains a VEV much smaller than the EW scale. This can be used in applications beyond the solution of the hierarchy problem. In particular, it can be used to suppress, but not remove, terms with particular symmetries. An example of where this is useful is neutrino physics, where the masses of the neutrinos are small but nonvanishing.  In this section we will work through this example in detail, restricting our attention to an effective field theory. 

As a first step we introduce a realization which is independent of the hierarchy problem. The model is exactly the same as in Section~\ref{sec:model} except, for now, we only assume that the total symmetry group of our model is $\mathbb{Z}_2$, i.e. we allow $D_8$ to be maximally broken. In this case $c_H\sim \mu^2/\Lambda^2$ and $\mu \ll \Lambda$ is achieved either through tuning or some other mechanism. In particular we do not assume that the VEV of $B$ is constrained by some cosmological selection mechanism. Here the VEV of $B$ must be small because the Higgs VEV is small.

 As with the type II seesaw mechanism (see e.g. \cite{Cai:2017mow} for a review), we introduce into our model a complex scalar field $\Delta\sim (\mathbf{1,3})_1$ (which we treat as a $2\times 2$ symmetric matrix). We suppose that it is odd under the $\mathbb{Z}_2$. Such a field does not permit a dimension-four interaction with fermions, but does permit a dimension-five one of the form 
\begin{align}\label{eq:neutrinomass}
\frac{y}{\Lambda} {L}^T \sigma_2\Delta\sigma_2 L B\,,
\end{align} 
where $L$ is the LH SM lepton doublet in the two-component Weyl convention, which is even under the $\mathbb{Z}_2$ and $y$ is an $\mathcal{O}(1)$ Wilson coefficient. 
The scalar potential in Section~\ref{sec:model} has to be extended with additional $\mathbb{Z}_2$-invariant interactions including $\Delta$
\begin{align} \label{eq:VLpotentail}
V_\Delta&=c_\Delta \Lambda^2 \mathrm{Tr}\Delta^* \Delta 
+c_{--} \Lambda \Phi_1^T \Delta^\ast \Phi_2\nonumber\\
&+\lambda_{++}B(\Phi_1^T\Delta^\ast \Phi_1+\Phi_2^T \Delta^\ast \Phi_2)+\lambda_{-+}B(\Phi_1^T \Delta^\ast \Phi_1-\Phi_2^T\Delta^\ast \Phi_2)\nonumber \\
&+\text{couplings with $\mathrm{Tr}\Delta^* \Delta$ }+ \mathrm{h.c.}.
\end{align}

Under the assumption that $V_\Delta$ does not dramatically affect the minimization of the potential in Section~\ref{subsec:PotentialMinimization}, we get a VEV in the neutral component of $\Delta$ which is dominated by the $c_{--}$ term and takes the form
\begin{align}
\langle \Delta \rangle \approx\frac{1 }{4c_\Delta\Lambda}c_{--} v^2 \sin 2\beta\sim \frac{\mu^2}{\Lambda} \sqrt{r}.
\end{align}
On substituting this into Eq.~\eqref{eq:neutrinomass} together with the VEV of $B$ we get
\begin{align} \label{eq:NuMasses}
m_\nu\sim \frac{\mu^4}{\Lambda^3}r .
\end{align}

The upper bound on neutrino masses from Planck data is $\sum m_{\nu_i}<0.12~\mathrm{eV}$~\cite{Planck:2018vyg}. In the standard type II seesaw model, which predicts a neutrino mass of $\sim v^2/\Lambda$, this would require new physics at the scale $\Lambda \sim10^{14}~\mathrm{GeV} $ for $v\sim 246~\mathrm{GeV}$. With $\mu\sim \frac{1}{2}m_h$, the model we have presented here can have a much lower scale of new physics: $\Lambda\sim 10^6~\mathrm{GeV}$ for $r=1$, and $\Lambda\sim 10^3~\mathrm{GeV}$ for $r=10^{-6}$.

However, if we assume an approximate $D_8$ symmetry which is only softly broken, and there exists some mechanism (like our solution to the hierarchy problem above) which makes $c_H$ small, then the neutrino masses gain an addition suppression. This is due to the fact that the $c_{--}$ term in Eq.~\eqref{eq:VLpotentail} and the dimension-five term in Eq.~\eqref{eq:neutrinomass} are not $D_8$ invariant. Thus both $y$ and $c_{--}$ have to contain powers of the soft-breaking parameter. Assuming that $y$ and $c_{--}$ scale like $y, c_{--}\sim \mu/\Lambda$ the neutrino masses are suppressed by an additional factor of $\mu^2/\Lambda^2$ compared to the expression in Eq.~\eqref{eq:NuMasses}. Note that the suppression depends on the UV completion and instead of $\mu$ also the fermion partner mass might appear. However, if we assume this particular scaling the expected cutoff $\Lambda$ ranges, with the same $\mu$ as above, from $10^4~\mathrm{GeV}$ at $r=1$ to  $10^3~\mathrm{GeV}$ at $r=10^{-6}$. As already mentioned this explanation for small neutrino masses can easily be combined with our solution to the hierarchy problem which for a unified explanation of small neutrino masses and a light Higgs predicts new physics at the TeV scale.

%%%%%%%%%%%%%%%%%%%% 
\section{Conclusion}
%%%%%%%%%%%%%%%%%%%%
%%
In this paper we presented a novel construction of a HVS operator, which can serve as a trigger operator in models which cosmologically select a low electroweak scale. The most compelling feature of our operator is that it is entirely made out of BSM degrees of freedom which are uncharged under the SM gauge group. This results in a reduced tuning in the hidden sector which cosmologically selects the electroweak scale.

Our model is based on a 2HDM extended by a real scalar field $B$ with a softly broken global $D_8$ symmetry (the symmetry group of a square). Due to the approximate $D_8$ symmetry the VEV of the real scalar $B$ tracks the Higgs VEV $\langle B \rangle \propto v^2/m_B$, such that $\mathcal{O}_{\rm HVS} =B^n$, $n\geq 1$ is the HVS operator with the desired properties. In order to ensure the approximate $D_8$ symmetry in the fermion sector we require vector-like fermionic partners for the SM fermions.

Paired with a hidden sector, such as the crunching sector of Ref.~\cite{Csaki:2020zqz}, which cosmologically selects small values of the $\langle B \rangle$, our model provides a compelling solution to the hierarchy problem. Some residual tuning of the order of $1\%$, however, is still required. This mainly corresponds to the little hierarchy between the Higgs mass and the mass of the vector-like fermion partners.

In a large part of parameter space the cosmological selection of the electroweak scale naturally pushes the 2HDM towards the alignment limit and favors a light CP-even Higgs scalar $s$ with mass $m_s \ll m_h$. The phenomenology of this light and weakly-coupled scalar together with further probes of our model were thoroughly discussed in Section~\ref{sec:pheno}.

While there are already various models which explain the electroweak scale through cosmological selection (see e.g.~\cite{Graham:2015cka,Arkani-Hamed:2016rle,Geller:2018xvz,Cheung:2018xnu,Giudice:2019iwl,Strumia:2020bdy,Csaki:2020zqz,Arkani-Hamed:2020yna,TitoDAgnolo:2021pjo,TitoDAgnolo:2021nhd}) our model has some unique features which we want to emphasize in the following. In contrast to previous realizations, the Higgs doublets in our model do not couple directly to the degrees of freedom which are responsible for the selection of the vacuum. This has the advantage that the Higgs itself neither mixes with degrees of freedom of the hidden sector nor does it have to be part of the hidden sector. Moreover, instead of being light and weakly-coupled the mediator between the Higgs and the hidden sector, i.e. the $B$ scalar in our model, is heavy and sits at the cutoff of the theory. 

In this setup the cosmological selection solves only half of the hierarchy problem. The other half of the solution is symmetry-based, for which the approximate $D_8$ symmetry is essential. Similarly to traditional solutions to the hierarchy problem, such as composite Higgs or little Higgs, the symmetry-based part requires new degrees of freedom at the TeV scale in the form of fermionic partners. However, in our model the mass scale of the fermionic partners is not directly related to the energy scale $\Lambda$ at which the full hierarchy problem is solved. Thus a discovery of fermion partners would not necessarily reveal the full mechanism behind the solution of the hierarchy problem.

The applications of our novel HVS operator are not limited to the hierarchy problem. In Section~\ref{sec:Neutrinos} we explored the possibility to obtain an additional suppression of neutrino masses in a variation of the type II seesaw mechanism using the smallness of the $B$ VEV. We showed that this allows us to lower the scale of new physics from roughly the GUT scale in the vanilla type II seesaw to the TeV scale in our model.

It is hoped that these two applications are not the only interesting ones for our model. It is further hoped that other ways to have SM-singlet HVS operators can be found, opening the possibility for distinct solutions to the applications above, and others as well.

\acknowledgments 
CC thanks the hospitality of the Aspen Center for Physics, which is supported by NSF grant PHY-1066293.
CC, AI, MR, and JTS are supported in part by the NSF grant PHY-2014071. CC is also supported in part by the BSF grant 2020220. AI is also supported in part by NSERC, funding reference number 557763. MR is also supported by a Feodor-Lynen Research Fellowship awarded by the Humboldt Foundation.

\appendix
%%%%%%%%%%%%%%%%%%%% 
\section{Details of the model}\label{sec:workingassumptions}
%%%%%%%%%%%%%%%%%%%%
 To aid in the clarity of the main text of the paper, certain specific details were excluded. This appendix will elucidate these details. 
 
 We start by rewriting the equations in Section~\ref{sec:model}, in the general case of $\lambda_1 \ne \lambda_2$, where $\lambda_{1,2} = \lambda_{1,2}'$. To do this, it is convenient to define the two parameters 
 \begin{align}
	a_{-1}=\frac{\lambda_1\lambda_2-\lambda_{345}^2}{\lambda_2^2-\lambda_{345}^2},\quad a_0= \frac{\lambda_1-\lambda_{345}}{\lambda_2-\lambda_{345}},
\end{align}
which are such that 
\begin{align}
\tan^2\beta&=\frac{2}{r}a_{-1}+a_0.
\end{align}
Here, $r$ is defined as in Eq.~\eqref{eq:rdefinition} except with $\lambda$ replaced with $\lambda_2$. The $\lambda_1=\lambda_2$ limit can be recovered by simply setting $a_{-1},a_0=1$.

The generic Higgs VEV is 
\begin{align}
\frac{v^2}{\mu^2}&=%2\frac{c_Hg^2k^2(-\lambda_2-\lambda_1+2\lambda_{345})+\mu^2(\lambda_1-\lambda_2)}{\lambda_1\lambda_2-\lambda_{345}^2},\\
%\frac{4}{\lambda_2-\lambda_{345}}(1+r)-\frac{2(\lambda_1-\lambda_2)}{\lambda_1\lambda_2-\lambda_{345}^2}r\\
\frac{4}{\lambda_2-\lambda_{345}}(1+r)+\frac{2}{\lambda_{345}}(1-\frac{a_{0}}{a_{-1}})r 
\end{align}
and the VEV of $B$ becomes 
\begin{align}
\frac{\Lambda}{\mu^2}\langle B \rangle=%\frac{2\sqrt{(c_Hg_\ast^2k^2(\lambda_2-\lambda_{345})+\mu^2(\lambda_2+\lambda_{345}))(c_Hg_\ast^2k^2(\lambda_1-\lambda_{345})-\mu^2(\lambda_1+\lambda_{345}))}}{\lambda_1\lambda_2-\lambda_{345}^2}.
%\frac{2}{\lambda_2-\lambda_{345}}\sqrt{\frac{2r}{\mathcal{N}_{-1}}+\frac{r^2}{\mathcal{N}_0}}.
-\frac{2c_{B\Phi}}{c_B(\lambda_2-\lambda_{345})}\sqrt{\frac{2r}{a_{-1}}+\frac{a_0r^2}{a_{-1}^2}}.
\end{align}

The physical parameters which are modified for $\lambda_1\ne \lambda_2$ are $\sin(\beta-\alpha)$ and $m_{h,s}$, which generalize to
\begin{align}
\sin(\beta-\alpha)&=-\frac{\lambda_2-\lambda_{345}}{\lambda_2}\sqrt{\frac{r}{2a_{-1}}}+\mathcal{O}\left(r^{3/2}\right),\\
	\frac{m_{h,s}^2}{\mu^2}&=\frac{2\lambda_2}{\lambda_2-\lambda_{345}}(1+r)+\frac{a_0-a_{-1}}{a_{-1}}r\nonumber\\&\pm \frac{2\lambda_2}{\lambda_2-\lambda_{345}}\sqrt{1 +\left(2\frac{\lambda_{345}^2}{\lambda_2^2}+\frac{\lambda_2-\lambda_{345}}{\lambda_{2}} \frac{a_0-a_{-1}}{a_{-1}}\right)r+\left(\frac{\lambda_{345}(a_0+1)}{2\lambda_2a_{-1}}\right)^2 r^2} .
\end{align}

We now turn to a discussion of the implicit assumptions made throughout the paper. The main mechanism of our model would not work, or would at least become more complicated, if other phases existed with $\langle B \rangle\ne 0$. In the region of parameter space with 
\begin{align}\label{eq:assump1}
	c_B,\quad - \lambda_{5}^\prime,\quad -\lambda_{45}^\prime, \quad \lambda_{1B}, \quad \lambda_{2B}\quad  > \quad 0
\end{align}
the only phase to exist with $\langle B \rangle\ne 0$ is the one described in the paper. Having $c_B>0$ prevents a phase existing with zero Higgs VEV. Having $ \lambda_{1B},  \lambda_{2B}>0$ prevents the existence of a phase with one Higgs VEV and $\langle B \rangle\ne 0$, whereas $ \lambda_{5}^\prime, \lambda_{45}^\prime >0$ prevents minima which break CP or electromagnetism.

There are natural assumptions which have to be made for the positivity of the potential. On top of those in Eq.~\eqref{eq:assump1} we need (at least) $\lambda_B>0$. Furthermore, to ensure the existence of our phase with $r>0$ we need 
\begin{align}
	 {\lambda_{1}^\prime}{}^2-\left({\lambda_{345}^\prime}-\frac{2 {c_{B\Phi}}{}^2}{{c_{B}}}\right){}^2>0.
\end{align}

 \sloppy There are certain conditions which guarantee that a small $\langle B \rangle$ implies small $v$. \emph{One set} of such conditions is given by ${\lambda_{2}^\prime}-{\lambda_{1}^\prime}\ge 0$, and the positivity of ${\lambda_{1}^\prime}-{\lambda_{345}^\prime}$ and $ {c_{B}} ({\lambda_{1}^\prime}-{\lambda_{345}^\prime}) ({\lambda_{2}^\prime}-{\lambda_{345}^\prime})
   ({\lambda_{1}^\prime}+{\lambda_{2}^\prime}-2 {\lambda_{345}^\prime})-{c_{B\Phi}}{}^2
   ({\lambda_{1}^\prime}-{\lambda_{2}^\prime}){}^2$. Notice that this last condition holds automatically if the separation of $\lambda_1^\prime$ and $\lambda_2^\prime$ is small.

Let us briefly discuss the feasibility of these assumptions. As long as $c_B$ is sufficiently larger than $2c_{B\Phi}^2$, then the conditions on primed parameters can be approximately translated to unprimed parameters. It can then be seen that nearly all our assumptions follow from $\lambda_{345}$ being small (see Eq.~\eqref{eq:lambda345Small}), and the necessary positivity conditions in Eq.~\eqref{eq:Positivity}, which generalize to 
\begin{align}
\lambda_{1,2}>0, \quad \lambda_3 > -\sqrt{\lambda_1\lambda_2}, \quad \lambda_{3}+\lambda_{4}-|\lambda_5|	> -\sqrt{\lambda_1\lambda_2}\,.
\end{align}

%%%%%%%%%%%%%%%%%%%% 
\section{Crunching mechanism}\label{sec:crunchingappendix}
%%%%%%%%%%%%%%%%%%%%
This appendix provides a detailed description of the crunching scenario introduced in Ref.~\cite{Csaki:2020zqz}, and how we can apply it to our model.

As explained in Section~\ref{sec:overview}, we imagine a multiverse of causally disconnected patches wherein a scanning sector sets the Higgs mass-squared parameter  in each patch, up to some cutoff scale $\Lambda$. To dynamically select a small Higgs VEV, we will introduce a spontaneously broken conformal sector that couples to a scalar singlet ``trigger'' operator $\mathcal{O}$. For our purposes we will take $\mathcal{O}$ to have mass dimension two and to be nonnegative. The dilaton $\chi$, a positive-valued singlet scalar field corresponding to the Goldstone boson of the broken scale invariance, mixes with the trigger operator. In the 5D dual description of the CFT, this means the fields which give rise to $\mathcal{O}$ must propagate in the AdS bulk.

The trigger operator must be sensitive to the Higgs VEV. The simplest choice is just $\mathcal{O} = \lvert H \rvert^2$, where $H$ here refers to the SM Higgs; this trigger was employed in Ref.~\cite{Csaki:2020zqz}. As we will shortly see, this choice leads to some undesirable phenomenology. For now we will discuss a general trigger operator, but eventually we will choose $\mathcal{O} = B^2$ in our model, where $B$ is the scalar singlet introduced in Section~\ref{sec:model}.

We introduce dynamics such that each Hubble patch rapidly undergoes a cosmological crunch unless the VEV of $\mathcal{O}$ is less than some critical value, $\langle \mathcal{O}  \rangle < \mathcal{O}_{\rm crit}$. This is possible because the dilaton potential is sensitive to the VEV of the trigger operator. We employ the Goldberger--Wise mechanism~\cite{Goldberger:1999uk} to generate a minimum in the dilaton potential, in which the vacuum energy is large and negative. Any patch in which the dilaton rolls down to this minimum will rapidly crunch. IR brane-localized interactions between $\mathcal{O}$ and the dilaton generate a second, metastable minimum in the potential, and this minimum may be long-lived on cosmological timescales. Crucially, the metastable minimum only exists for $\langle \mathcal{O}  \rangle < \mathcal{O}_{\rm crit}$.

The result of the crunching dynamics is that the only patches of the multiverse which survive until the present day are those in which $\langle \mathcal{O} \rangle < \mathcal{O}_{\rm crit}$. In these patches, the dilaton can safely live in the metastable minimum, and the cosmological history is conventional. All other patches roll down to the true vacuum and crunch. The value of the Higgs VEV $v$ corresponding to the critical value of the trigger operator is hierarchically smaller than the cutoff $\Lambda_H$, leading to what appears to be a naturalness problem.

%%%%%%%%%%%%%%%%%%%%%%%%%%%%%%%%%%%
\subsection{Dilaton potential}
%%%%%%%%%%%%%%%%%%%%%%%%%%%%%%%%%%%

More concretely, the dilaton potential is given by
\begin{equation}\label{eq:dilatonpotential}
    V(\chi, \mathcal{O}) = V_{\rm GW}(\chi) + V_{\mathcal{O}\chi}(\chi, \mathcal{O}) ,
\end{equation}
where
\begin{equation}\begin{split}
    V_{\rm GW} &=  -\kappa \chi^4 + \kappa_{\rm GW} \frac{\chi^{4+\delta}}{k^\delta}, \\
    V_{\mathcal{O}\chi} &= \kappa_2 \mathcal{O} \frac{\chi^{2+\gamma}}{k^\gamma} - \kappa_\epsilon \mathcal{O} \frac{\chi^{2+\gamma+\epsilon}}{k^{\gamma+\epsilon}} - \kappa_4 \mathcal{O}^2 \frac{\chi^{2\gamma}}{k^{2\gamma}} .
\end{split}\end{equation}
Here $k$ is the inverse AdS curvature, which would be identified with the UV cutoff $\Lambda$ of the theory. The terms in $V_{\rm GW}$ arise from the usual Goldberger--Wise mechanism. There is the scale-invariant quartic term and the $\chi^{4+\delta}$ term which corresponds to a small explicit breaking of scale invariance.
The mixing terms in $V_{\mathcal{O}\chi}$ come from an IR brane-localized potential for $\mathcal{O}$. The $\kappa_2$ and $\kappa_4$ terms arise respectively from brane-localized $\mathcal{O}$ and $\mathcal{O}^2$ terms. When we take $\mathcal{O} = B^2$, these correspond to brane-localized quadratic and quartic terms in $B$. The parameter $\gamma$ is related to the bulk scaling of $\mathcal{O}$: $\mathcal{O} \sim z^{1 - \gamma/2}$. Lastly, allowing terms involving a field with an approximately marginal dimension $\epsilon$, such as the Goldberger--Wise scalar, yields the $\kappa_\epsilon$ term. Since we assume $\mathcal{O}$ has mass dimension two, these are the only renormalizable terms allowed in the potential.

\begin{figure}
    \centering
    \includegraphics[width=0.7\textwidth]{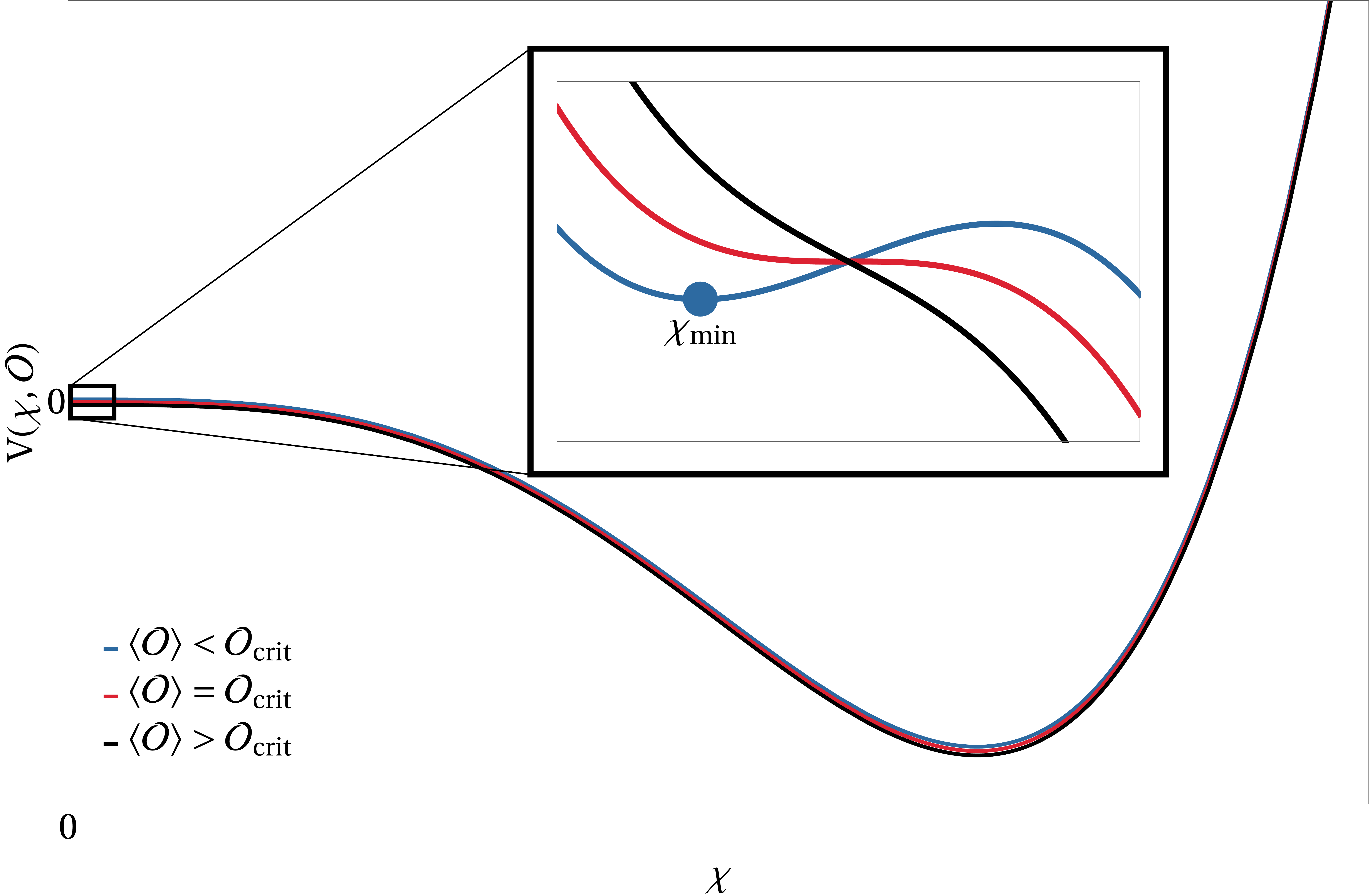}
    \caption{Sketch of the dilaton potential in Eq.~\eqref{eq:dilatonpotential}, adapated from Ref.~\cite{Csaki:2020zqz}. When the VEV of the trigger operator $\mathcal{O}$ is less than $\mathcal{O}_{\rm crit}$, there is a metastable minimum at $\chi = \chi_{\rm min}$ (inset, blue curve). The metastable minimum disappears as $\langle \mathcal{O} \rangle$ is raised beyond $\mathcal{O}_{\rm crit}$ (red and black curves), leaving only the true vacuum, which has a large negative energy density.}
    \label{fig:crunchingpotential}
\end{figure}

A sketch of the potential is shown in Figure~\ref{fig:crunchingpotential}. This illustrates the existence of a second metastable minimum in the potential at $\chi = \chi_{\rm min}$, which disappears as the VEV of $\mathcal{O}$ is increased beyond the critical value. One can estimate
\begin{equation}
    \chi_{\rm min} \sim \langle \mathcal{O} \rangle \sim k \left( \frac{\kappa_2}{\kappa_{\epsilon}} \right)^{1/\epsilon} .
\end{equation}
Thus, a mild hierarchy between $\kappa_2$ and $\kappa_{\epsilon}$ can generate a large hierarchy $\langle \mathcal{O} \rangle, \chi_{\rm min} \ll k$, thanks to the conformal symmetry.

In order to fully solve the hierarchy problem, one must also introduce a mechanism to forbid vanishing Higgs VEVs (corresponding to positive Higgs mass-squared parameter). There are multiple ways to accomplish this. One way is to introduce a confining gauge group in the AdS bulk. This generates an explicit breaking of scale invariance at small $\chi$, which adds a term to the dilaton potential of the form $\chi^\alpha \tilde{\Lambda}^{4 - \alpha}$, where $\tilde{\Lambda}$ is the confining scale. The effect of this term is to generate a minimum VEV $\mathcal{O}_\emptyset$, such that all patches where $\langle \mathcal{O} \rangle < \mathcal{O}_\emptyset$ will crunch. (We assume that $\langle \mathcal{O} \rangle$ is small or vanishing when $v= 0$.)

Another option is to use self-organized localization~\cite{Giudice:2021viw} to disfavor a small or vanishing Higgs VEV. In this approach, the potential of the scanning sector causes patches with larger $\langle \mathcal{O} \rangle$ (but still less than $\mathcal{O}_{\rm crit}$) to inflate more rapidly. Consequently, the multiverse is dominated by patches in which the VEV of $\mathcal{O}$ is very close to $\mathcal{O}_{\rm crit}$.

%%%%%%%%%%%%%%%%%%%%%%%%%%%%%%%%%%%
\subsection{Trigger operators}
%%%%%%%%%%%%%%%%%%%%%%%%%%%%%%%%%%%

As stated above, the simplest choice of trigger operator is $\lvert H \rvert^2$. However, this requires the Higgs to propagate in the bulk, and therefore the electroweak gauge bosons must live in the bulk as well. The model then includes KK modes of the $W$ and $Z$, whose masses are set by the location of the metastable minimum $\chi_{\rm min}$. Experiments constrain these KK partners to lie at the TeV scale or higher. To avoid these constraints, we must have $\chi_{\rm min} \gtrsim 1$~TeV, which introduces some fine-tuning into the model.

Here we instead choose the trigger $B^2$. Hence, $B$ propagates in the bulk while all the other particles lie on the UV brane. This is possible because $B$ is an SM singlet.
In order to solve the hierarchy problem, the critical value of the $B$ VEV at which crunching occurs, $B_{\rm crit}$, must obey
\begin{equation}
    B_{\rm crit} \lesssim \frac{v^2}{k} \Leftrightarrow r_{\rm crit}\lesssim 1 .
\end{equation}

%%%%%%%%%%%%%%%%%%%%%%%%%%%%%%%%%%%
\subsection{Phenomenology and cosmology}
%%%%%%%%%%%%%%%%%%%%%%%%%%%%%%%%%%%
We now consider the possible phenomenological and cosmological ramifications of using the crunching mechanism in our model. Since the SM particles are localized on the UV brane, they couple very weakly to IR-localized modes. The KK modes of the $B$ as well as the dilaton are IR-localized, so they are essentially irrelevant for phenomenology.

The would-be zero mode of the $B$ gets its potential partly on the UV brane and partly on the IR brane. The UV brane-localized potential for $B$ causes the would-be zero mode to get a large mass $m_B^2 =  1/2 c_B \Lambda^2\sim k^2$ (see eq.~\eqref{eq:VB}). Since the cutoff scale $k$ lies far above the electroweak scale, this mode is not observable at colliders.

Employing the crunching mechanism places cosmological constraints on the model. For the dilaton potential to be sensitive to VEVs of order $B_{\rm crit} \lesssim v^2/k$, the Hubble scale during inflation must be less than $B_{\rm crit}$. The corresponding bound on the scale of inflation $M_I$ is
\begin{equation}
    M_I \lesssim \sqrt{ \frac{M_P}{k} } v .
\end{equation}
where $M_P$ is the reduced Planck mass.

Also, we must ensure that the total vacuum energy density in the true vacuum of the theory is always negative, so that a cosmological crunch is triggered by the dilaton rolling down to the global minimum of its potential. We therefore require $k > M_I$, so that the dilaton potential in the true vacuum, which is of order $-k^4$, dominates over any contribution to the vacuum energy from the inflaton sector. Combining this with the upper bound on $M_I$, it is easy to see that $k > M_I$ is always satisfied for $k \gtrsim v^{2/3} M_P^{1/3} \sim 10^4$~TeV.

Assuming the universe is radiation-dominated immediately after reheating, the Hubble constant at reheating satisfies
\begin{equation}
    H = \left( \frac{g_* \pi^2}{90} \right)^{1/2} \frac{T_{\rm RH}^2}{M_P}
\end{equation}
where $T_{\rm RH}$ is the reheating temperature. This leads to an upper bound
\begin{equation}
    T_{\rm RH} \lesssim \left( \frac{90}{g_* \pi^2} \right)^{1/4} \sqrt{ \frac{M_P}{k} } v .
\end{equation}

Lastly, if a dark confining gauge group in the bulk is used to crunch away patches with $\langle B \rangle < B_{\emptyset}$, we clearly must require $B_{\emptyset} < B_{\rm crit}$. The dynamical scale of the gauge group $\tilde{\Lambda}$ sets the scale of $B_{\emptyset}$, and therefore we have
\begin{equation}
    \tilde{\Lambda} \lesssim \frac{v^2}{k} .
\end{equation}
\bibliographystyle{JHEP.bst}
\bibliography{draft}
\end{document}